\documentclass[twocolumn]{aastex62}
\usepackage{amsmath}   
\usepackage{wasysym}
\usepackage{graphicx}
\usepackage{amssymb}
\usepackage{epstopdf}
\usepackage{mathrsfs}
\usepackage{anyfontsize}
\usepackage{natbib}
\usepackage{color}
\usepackage{lipsum}
\usepackage{diagbox}

\shorttitle{Low-energy explosions in a gravitational field}
\shortauthors{Paradiso, Coughlin, Zrake, \& Pasham}
\begin{document}
\title{Low-energy Explosions in a Gravitational Field: Implications for Sub-energetic Supernovae and Fast X-ray Transients}
\author[0009-0003-8285-0702]{Daniel A.~Paradiso}
\affiliation{Department of Physics, Syracuse University, Syracuse, NY 13210, USA}
\author[0000-0003-3765-6401]{Eric R.~Coughlin}
\affiliation{Department of Physics, Syracuse University, Syracuse, NY 13210, USA}
\author[0000-0002-1895-6516]{Jonathan Zrake}
\affiliation{Clemson University, 118 Kinard Laboratory Clemson, SC 29634, USA}
\author[0000-0003-1386-7861]{Dheeraj R.~Pasham}
\affiliation{MIT Kavli Institute for Astrophysics and Space Research, Massachusetts Institute of Technology, Cambridge, MA 02139, USA}

\email{dparadis@syr.edu}
\email{ecoughli@syr.edu}

\begin{abstract}
Observations and theory suggest that core-collapse supernovae can span a range of explosion energies, and when sub-energetic, the shockwave initiating the explosion can decelerate to speeds comparable to the escape speed of the progenitor. In these cases, gravity will complicate the explosion hydrodynamics and conceivably cause the shock to stall at large radii within the progenitor star. To understand these unique properties of weak explosions, we develop a perturbative approach for modeling the propagation of an initially strong shock into a time-steady, infalling medium in the gravitational field of a compact object. This method writes the shock position and the post-shock velocity, density, and pressure as series solutions in the (time-dependent) ratio of the freefall speed to the shock speed, and predicts that the shock stalls within the progenitor if the explosion energy is below a critical value. We show that our model agrees very well with hydrodynamic simulations, and accurately predicts (e.g.) the time-dependent shock position and velocity and the radius at which the shock stalls. Our results have implications for black hole formation and the newly detected class of fast X-ray transients (FXTs). In particular, we propose that a ``phantom shock breakout'' -- where the outer edge of the star falls through a stalled shock -- can yield a burst of X-rays without a subsequent optical/UV signature, similar to FXTs. This model predicts the rise time of the X-ray burst, $t_{\rm d}$, and the mean photon energy, $kT$, are anti-correlated, approximately as $T \propto t_{\rm d}^{-5/8}$. 
\end{abstract}


\section{Introduction}
\label{sec:intro}
Most massive stars end their lives in spectacular fashion in what are known as core-collapse supernovae (CCSN), which are initiated when the star runs out of nuclear fuel in its core, causing the core to collapse under its own self-gravity. During the core-collapse, electron capture causes the core to become increasingly destabilized due to the diminishing support from electron degeneracy pressure and at the same time produces an abundance of neutrons. The byproduct of this neutron rich and electron deficient environment is a proto-neutron star (PNS). As the density of the PNS increases, the equation of state stiffens due to the repulsive strong nuclear force and neutron degeneracy pressure. This causes the PNS to ``bounce,'' creating an outward propagating shockwave. If the shockwave is energetic enough, it will propagate through and unbind the overlying stellar envelope in a CCSN \citep{Arnett66}. However, while the fiducial energy of the ejected mass in a CCSN is on the order of $\sim10^{51}$\,erg, there is both observational and theoretical evidence that not all CCSNe are highly energetic or even successful (e.g., \citealt{horiuchi11, smartt15, Adams17b, Kuroda22}). Multiple CCSNe are observed on a nightly basis by time-domain surveys such as the Zwicky Transient Facility \citep{bellm17}, the All-Sky Automated Search for Supernovae \citep{shappee14}, and -- imminently -- the Rubin Observatory/LSST \citep{Ivezic19}. It is therefore important to understand the physical conditions that lead to failed, nearly failed, and successful CCSNe.

During the initial propagation of the shockwave that follows the PNS bounce, it encounters an infalling ambient medium and loses energy through the dissociation of heavy nuclei, and eventually stalls at a radius $R\sim 100 - 200$\,km (\citealt{Bethe90} and references therein). However, the formation of the PNS is accompanied by the release of $\sim10^{53}$\, ergs of gravitational binding energy in the form of neutrinos. During this radiation of mass-energy, neutrino annihilations deposit energy in the post-shock region, which heats the region behind the shock and may drive a successful explosion \citep{Colgate66, Bethe85}. Without such a revival mechanism, the shock will remain stalled and material will continue to accrete onto the neutron star until a black hole is formed. 

The transition between the stalled accretion shock phase and explosion in these neutrino-powered supernovae occurs when the neutrino luminosity of the PNS exceeds a critical value. This critical neutrino luminosity, $L^{\rm crit}_{\nu,\rm core}$, was initially calculated by \cite{Burrows93}, who showed that for a fixed mass accretion rate, there exists a PNS neutrino luminosity such that no steady-state solution exists in the region between the PNS and the stalled accretion shock. They then argued that any PNS neutrino luminosity above this critical value will result in the revival of the stalled shock and lead to a successful supernova. Using this critical luminosity along with the results obtained by \cite{Duncan} -- who studied neutrino driven winds of neutron stars -- \cite{Burrows93} also found an estimate for supernova energies. Their results showed that the total mechanical energy imparted into the supernova after the blast, $E_{\rm s}$, is $\propto {L_{\nu_e}}^{3.5}$, which implies that a wide range of supernova energies can exist due to the steep power law in $L_{\nu_e}$. 

The physics behind the existence of $L^{\rm crit}_{\nu,\rm core}$ has since been studied to better understand both failed and successful CCSNe. For example, \cite{Pejcha} showed that an ``antesonic'' condition exists, such that $L^{\rm crit}_{\nu,\rm core}$ corresponds to a constant ratio of the sound speed to the escape velocity in the accretion flow. \citet{Gogilashvili22} derived a ``force explosion condition'' for a successful explosion and from which they obtained the critical neutrino luminosity and antesonic conditions, thus uniting the two conditions. In a follow-up paper, they also showed that their condition is consistent with one-dimensional (1D), spherically symmetric simulations (\citealt{Gogilashvili23}). Three-dimensional simulations of CCSNe have recently been performed (see \citealt{Muller17,Oconnor18, Summa18, burrows2019three, Vartanyan19}), which demonstrate that the incorporation of neutrino heating can lead to a successful explosion in some cases. 

While the revival mechanism for successful explosions is still uncertain, the eventual escape of the shockwave from the star -- should it be revived by neutrino heating or other effects -- is constrained by the energy behind the blast (following the deposition of energy and momentum by neutrinos) and the properties of the overlying stellar envelope. This aspect of the ``explodability'' of the star can be understood through an analysis of the propagation of the shockwave and the post-shock gas, and self-similarity --  which reduces the Euler (or Navier-Stokes) equations to a set of ordinary differential equations that can be solved straightforwardly and numerically (or, in special cases, analytically) -- is among the most useful methods in this regard.
Perhaps the most well-known self-similar solution to the Euler equations is the Sedov-Taylor blastwave \citep{Sedov456, Taylor50}, which can be used to describe the propagation of the shock as well as the post-shock fluid variables when the supernova energy is large compared to the binding energy of the star. However, this inequality is not always satisfied, and there are scenarios in which the energy of the blast is comparable to the binding energy of the star. In such cases, the gravitational field and infalling material will be relevant in determining the dynamics of the shock, making the Sedov-Taylor blastwave inapplicable. However, it was recently shown \citep{CQR1, CQR2, CQR3, coughlin23} that there is a distinct class of self-similar, weak-shock solution to the fluid equations, in which the Mach number is only marginally greater than one. These solutions also result in fallback accretion onto the compact object, and are relevant to completely failed supernovae where the shock is always sub-energetic (and, in this case, the shock is generated through the mass lost to neutrinos and the hydrodynamical response of the overlying envelope; \citealt{Nadezhin80}). 

Here we focus on the bridge between these two regimes. That is, we study the transition of a shock with an initial velocity much greater than the freefall speed into one with a velocity comparable to the freefall speed due to the presence of a gravitational field, ultimately to understand low-energy supernova explosions\footnote{\citet{Yalinewich21} considered a similar setting, but for a medium with an initial density profile $\rho \propto r^{-2}$, for which the Sedov-Taylor shock velocity also scales as the free-fall speed. As such, their solutions are still self-similar, whereas our work investigates the more general scenario in which self-similarity is violated owing to the difference in scaling between the strong-shock velocity (with radius) and the freefall speed.}. In Section~\ref{sec:estimates} we provide an analytical estimate for the minimum energy of a successful supernova by assuming that the explosion is initially strong, such that the Sedov-Taylor scaling for the shock velocity is initially upheld, and equating this velocity to the infall speed. We show that when the density profile of the ambient medium scales as $\propto r^{-3/2}$, this estimate also results from equating the explosion energy to the binding energy of the overlying envelope, but for other power-law density profiles these two methods yield distinct energy estimates. In Section~\ref{sec:Analysis} we analytically solve the fluid equations, which account for the gravitational field of a point mass, through a series expansion of the shock position and fluid variables in the initially small (and time-dependent) ratio of the freefall speed to the shock speed. Therefore, for a shock that is initially strong, these perturbations account for small corrections to the unperturbed, Sedov-Taylor solution but will become increasingly important as the shock decelerates and its speed becomes comparable to the freefall speed. This solution predicts that the shock will stall after propagating a finite distance, which we use to derive a minimum energy condition for a successful explosion. In Section~\ref{sec:simulations} we compare our analytical model with 1D hydrodynamical simulations, performed with {\sc flash} \citep{fryxell00}. We show that our method is able to very accurately predict the time-dependent propagation of the shock and the radius at which the shock stalls, as well as the post-shock fluid velocity, density, and pressure profiles. 

In Section~\ref{sec:discussion} we discuss the astrophysical and observational implications of our findings as well as caveats of our work. In particular, we discuss black hole formation, as well as the possible generation of a unique astrophysical transient generated by the passing of the outer envelope of a star passing through a stalled shock. We show that the observational signatures of such an event, which we term a ``phantom shock breakout,'' are consistent with the newly observed class of fast X-ray transients, which are short flashes of soft X-rays of as-yet-unknown origin. We summarize and conclude in Section~\ref{sec:conclusion}.

\section{Analytic Estimates} 
\label{sec:estimates}
After the shock stalls and is revived, it reaches a radius where it is no longer accelerating due to the energy provided by neutrinos and enters an energy-conserving 
regime. We characterize this radius by an ambient density $\rho_{\rm i}$ and length scale $R_{\rm i}$. We initially assume that the medium into which the shock is propagating has a density profile  $\rho \propto r^{-3/2}$ and that the gas is in time-steady and pressureless freefall. It therefore follows that the total mass swept up by the shockwave is
\begin{equation}
    M \simeq \rho R^3 \simeq \rho_{\rm i}R_{\rm i}^3\left(\frac{R}{R_{\rm i}}\right)^{3/2}. 
\end{equation}
If the energy $E \simeq MV^2$ is conserved, then the shock speed varies as a function of shock position as
\begin{equation}
    V=\sqrt{\frac{E}{\rho_{\rm i}{R_{\rm i}}^3}}\left(\frac{R}{R_{\rm i}}\right)^{-3/4} \equiv V_{\rm i}\left(\frac{R}{R_{\rm i}}\right)^{-3/4},
\end{equation}
where $V_{\rm i} \equiv \sqrt{E/\left(\rho_{\rm i}R_{\rm i}^3\right)}$. The infalling material, on the other hand, has a speed $v = \sqrt{2GM_{\rm ns}/R}$ where $M_{\rm ns}$ is the mass of the neutron star. The shock velocity therefore falls off at a faster rate than that of the infalling material, and there will be a point when the velocity of the infalling material is greater than the velocity of the shock. At this distance the shock will stall under the ram pressure of the overyling gas, and equating the shock speed to the freefall speed shows that this radius, which we denote $R_{\rm s}$, is
\begin{align}
\begin{split}
   & V_{\rm i}\left(\frac{R_{\rm s}}{R_{\rm i}}\right)^{-3/4} = \sqrt{\frac{2GM}{R_{\rm s}}} \\
   & \Rightarrow \quad R_{\rm s} = R_{\rm i}\left(\frac{V_{\rm i}}{\sqrt{2GM_{\rm ns}/R_{\rm i}}}\right)^{4} \equiv R_{\rm i}\epsilon_{\rm i}^{-4}. \label{Rsofeiapp}
    \end{split}
\end{align}
Here we defined
\begin{equation}
    \epsilon_{\rm i} = \frac{1}{V_{\rm i}}\sqrt{\frac{2GM}{R_{\rm i}}}
\end{equation}
as the ratio of the freefall speed to the shock speed, which, if the shock is initially strong and in the Sedov-Taylor regime, should be much less than one. Equation \eqref{Rsofeiapp} shows that the stall radius is a very sensitive function of this ratio, which arises from the fact that the Sedov-Taylor velocity falls off only slightly faster than the freefall speed. Therefore, small changes in the initial velocity of the shock correspond to large changes in the location at which it stalls, e.g., a change in $\epsilon_{\rm i}$ by a factor of 2 result in a change in the stall radius by over an order of magnitude.

We can also use our definition of $V_{\rm i}$ in terms of the energy to write the stall radius in terms of the initial explosion energy; doing so gives
\begin{align}
\begin{split}
    & \sqrt{\frac{E}{\rho_{\rm i}{R_{\rm i}}^3}}\left(\frac{R}{R_{\rm i}}\right)^{-3/4} = \sqrt{\frac{2GM}{R}} \\
    & \Rightarrow \quad 
    \frac{R_{\rm s}}{R_{\rm i}} = \left(\frac{E}{ GM_{\rm ns}\rho_{\rm i}{R_{\rm i}}^2} \right)^2 .\label{stall radius}
    \end{split}
\end{align}
A successful explosion will occur if $R_{\rm s} \geq R_{\star}$, where $R_{\star}$ is the radius of the star. Setting $R_{\rm s} = R_{\star}$ then yields a lower limit on the energy that will yield a successful explosion:
\begin{equation}
   E \gtrsim GM_{\rm ns}\rho_{\rm i}{R_{\rm i}}^2\left(\frac{R_{\star}}{R_{\rm i}}\right)^{1/2} \label{energyapprox} 
\end{equation} 
where, using fiducial values associated with supergiant progenitors (see Figure \ref{fig:temp}) of $M_{\rm ns}=1.5$\,\(M_\odot\), $\rho_{\rm i}=10^{-5}~\rm g~cm^{-3}$, ${R_{\rm i}}=10^{12}$\,cm and $R_{\star}=10^{14}$\,cm, we obtain an estimate for $E$ that is of the order $10^{46}$\,ergs.

The preceding analysis let the ambient medium be in time-steady freefall, which is only a self-consistent assumption if the ambient gas satisfies $\rho(r) \propto r^{-3/2}$. This may be a good approximation over certain ranges in radii of a massive star where the gas is convective, ideal and monatomic gas-pressure dominated, and the enclosed mass is not strongly growing with radius, as in this case $p \propto \rho^{5/3}$ and -- from the equation of hydrostatic balance -- $\rho \propto r^{-3/2}$. In general, however, the density profile of the progenitor of a core-collapse supernova will be more complicated, not least because of the succession of nuclear burning shells throughout the interior (see Figure \ref{fig:temp} below for examples). The ambient density at the location of the shock will also be time-dependent, not just because of the variable ambient density throughout the star, but also because a rarefaction wave is traveling through the overlying envelope causing the shells to collapse to the center (i.e., the entire star is not instantaneously in freefall at all radii).  If, however, the ambient medium is at least well-approximated by a power-law density profile of the form $\rho \propto r^{-n}$, which is generally valid for the outer envelopes of red supergiants owing to their convective envelopes and shallow mass profile (see, e.g., Figures 7 \& 12 of \citealt{coughlin18}), then there is a self-similar solution for the propagation of the rarefaction wave traveling into the envelope and the $\sim$ freefalling material behind the shock \citep{coughlin19}. In this case, the density profile at small radii within the star is 
\begin{equation}
    \rho \propto t^{1-2n/3}r^{-3/2}.
\end{equation}
If the shock is strong, then the energy is conserved, and we have\footnote{It is interesting to note that Equation \eqref{Roft} is nearly identical to the Sedov-Taylor scaling -- being $R\propto t^{2/(5-n)}$ -- for $3/2 \le n \le 2$ (note that the two are equal for $n = 3/2$ and $n = 2$; the maximum difference between the two and within these limits is $\simeq 0.0037$ at $n \simeq 1.76$). However, as $n$ exceeds 2, the Sedov-Taylor scaling is significantly steeper. As such, the shock would not reach the accelerating regime, in which the temporal power-law index is greater than 1, until $n > 3.75$. This power-law index can be contrasted with the value of $n$ above which the Sedov-Taylor blastwave accelerates, being $n = 3$, showing that the ambient medium must fall off exceptionally steeply with radius for the shock to enter the accelerating regime.}
\begin{equation}
    E \sim \rho R^3 V^2 \quad \Rightarrow \quad R \sim t^{\frac{2}{7}\left(1+2n/3\right)}. \label{Roft}
\end{equation}

\begin{figure}
    \includegraphics[width=0.47\textwidth]{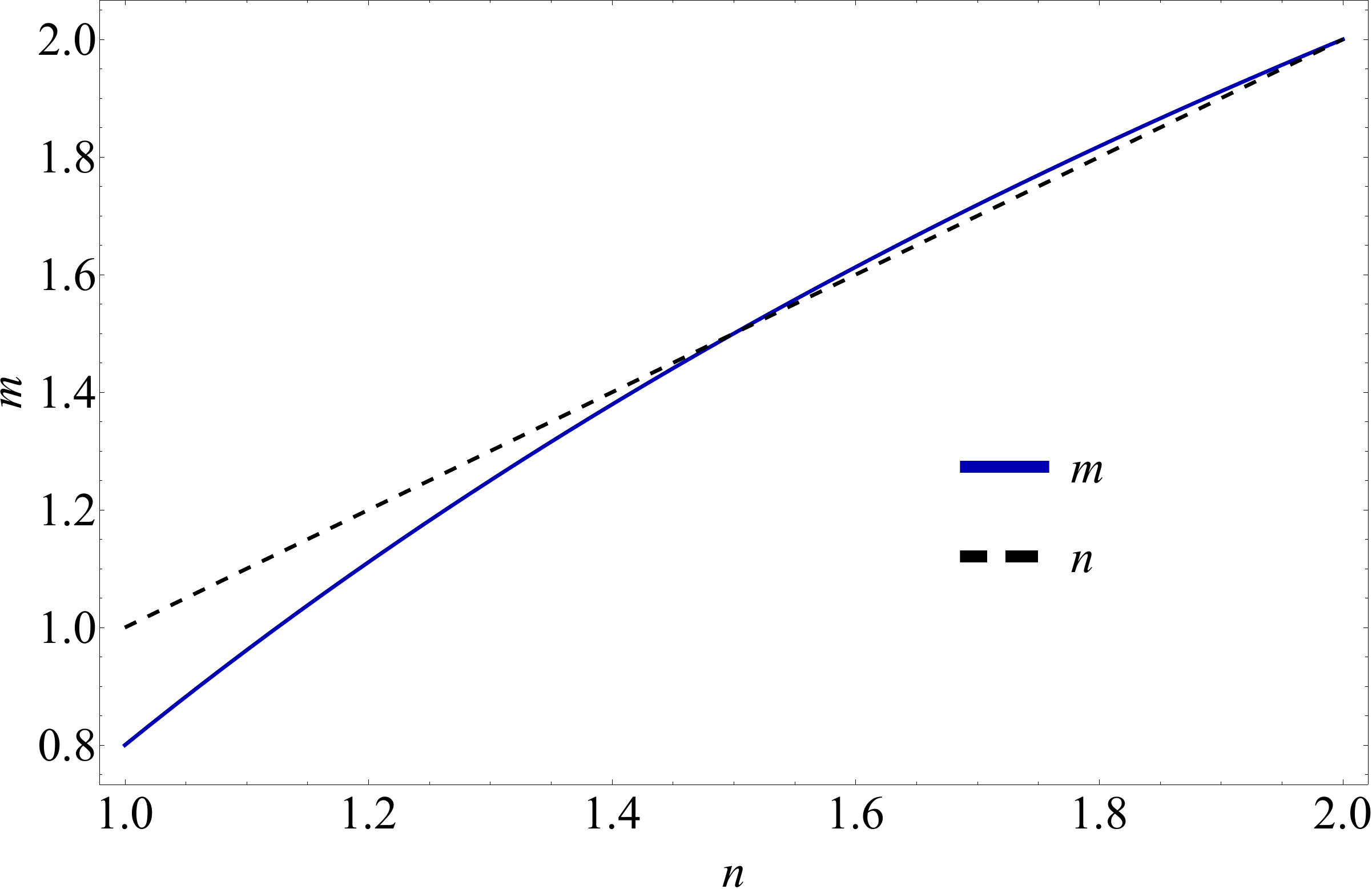}
    \caption{The effective power-law index of the ambient gas during freefall at small radii, i.e., if the stellar envelope at large radii has a density profile $\propto r^{-n}$, then $\rho \propto r^{-m}$ with $m = (10n-6)/(2n+3)$ at the location of the shock.}
    \label{fig:m_of_n}
\end{figure}

Inverting Equation \eqref{Roft} and solving for $t(R)$, we see that the density at the location of the shock satisfies
\begin{equation}
    \rho(R) \propto R^{-\frac{10n-6}{2n+3}} \equiv R^{-m}, \label{rhom}
\end{equation}
which is just a power-law profile with an \emph{effective} power-law index of $m = (10n-6)/(2n+3)$. Figure \ref{fig:m_of_n} shows $m$ as a function of $n$ by the blue curve, and $n$ (for reference) by the black, dashed line. We see that the two are equal for $n = 3/2$ and $n = 2$, but $m$ is slightly larger than $n$ for $3/2<n<2$, and $m$ is smaller than $n$ for $n < 3/2$. This difference implies that the density profile is effectively shallower than the ambient density if $n < 3/2$. 

If the shock is initially strong, then as the shock propagates outward into the infalling gas we have
\begin{equation}
    \rho R^3V^2 = E \,\,\, \Rightarrow \,\,\, \Rightarrow V \simeq \sqrt{\frac{E}{\rho_{\rm i}R_{\rm i}^3}}\left(\frac{R}{R_{\rm i}}\right)^{\frac{m-3}{2}},
\end{equation}
and equating this to the freefall speed, setting the radius at which this occurs as the stellar radius, and solving for the energy gives
\begin{equation}
    E \gtrsim G M\rho_{\rm i}R_{\rm i}^2\left(\frac{R_{\star}}{R_{\rm i}}\right)^{\frac{6\left(2-n\right)}{3+2n}}. \label{Egen}
\end{equation}
Equation \eqref{Egen} yields Equation \eqref{energyapprox} when $n = 3/2$, but because in general $m \neq n$ in Equation \eqref{rhom}, it is not true that Equation \eqref{Egen} yields the same condition that would result from requiring that the initial shock energy equal the binding energy of the envelope with density profile $\propto r^{-n}$. In particular, the binding energy of the envelope with density profile $\propto r^{-n}$ is $\propto \left(R_{\star}/R_{\rm i}\right)^{2-n}$. Therefore, if $n < 3/2$, Equation \eqref{Egen} is \emph{more restrictive} on the energy of the explosion, meaning that it is more difficult to create a successful explosion than would be inferred from simple energy conservation; this is consistent with the fact that $m < n$ if $n < 3/2$, as can be seen from Equation \eqref{rhom}.

The analysis leading to Equations \eqref{energyapprox} \& \eqref{Egen} is approximate from the standpoint that we assumed that the shock will maintain the Sedov scaling up until $R_{\rm s}$. In actuality, however, the shock will begin to decelerate more rapidly as the effects of the gravitational field become more pronounced, and hence the stall radius could be substantially smaller than these estimates suggest. In the next section we use a series expansion approach to rigorously derive the shock position as a function of time, as well as the spatial and temporal evolution of the post-shock fluid quantities, as it evolves in the gravitational field of a compact object and propagates through an infalling, $\rho \propto r^{-3/2}$ medium.

\section{Series solution}
\label{sec:Analysis}
We assume that the explosion is spherically symmetric and creates a shockwave that is initially highly energetic, propagating into a freefalling ambient medium with a $\propto r^{-3/2}$ density profile, and subject to the gravitational field of a central compact object. Accounting for the gravitational field of the neutron star with mass $M_{\rm ns}$, the spherically symmetric continuity, radial momentum and entropy equations are given by
\begin{align}
    &\frac{\partial{\rho}}{\partial{t}}+\frac{1}{r^2}\frac{\partial{}}{\partial{r}}\left[\rho r^2v\right]=0, \label{continuity}\\ 
    &\frac{\partial{v}}{\partial{t}}+v\frac{\partial{v}}{\partial{r}}+\frac{1}{\rho}\frac{\partial{p}}{\partial{r}}=-\frac{GM_{\rm ns}}{r^2}, \label{momentum}\\ 
    &\frac{\partial{s}}{\partial{t}}+v\frac{\partial{s}}{\partial{r}}=0. \label{energy eq}
\end{align}
Here $r$ is spherical radius, $v$ is the radial velocity, $\rho$ is the density, $p$ is the pressure, $s=\ln\left(p/\rho^\gamma\right)$ is the specific entropy and $\gamma$ is the adiabatic index of the post-shock gas. We define the time-dependent position of the shock as $R(t)$ and the shock velocity as $V(t) = dR/dt$. The relevant time and length scales are therefore determined by the shock position and velocity. 

Equations~\eqref{continuity}-\eqref{energy eq} describe the evolution of the fluid behind the shock front, at which point the mass, momentum, and energy fluxes are continuous. The shock jump conditions yield a set of boundary conditions for the post-shock velocity, density, and pressure that guarantee the continuity of these fluxes across the shock, and these conditions are
\begin{align}
    v\left(R\right) &=\left[\frac{2}{\gamma+1}-\frac{1}{V}\sqrt{\frac{2GM_{\rm ns}}{R}}\frac{\gamma-1}{\gamma+1}\right] V, \label{v bc} \\
    \rho\left(R\right)&=\frac{\gamma+1}{\gamma-1}\rho_{\rm i}{\left(\frac{R}{R_{\rm i}}\right)}^{-3/2}, \label{rho bc} \\
    p\left(R\right) &= \left(1+\frac{1}{V}\sqrt{\frac{2GM_{\rm ns}}{R}}\right)^2\frac{2}{\gamma+1}\rho_{\rm i}{\left(\frac{R}{R_{\rm i}}\right)}^{-3/2}V^2, \label{p bc}
\end{align}
where, $\rho_{\rm i}$ is the ambient density at the location of the shock. 

When there is no gravitational field (i.e., for $M \equiv 0$), the solution is given by the Sedov-Taylor blastwave\footnote{Assuming that the initial conditions are given by the Sedov-Taylor blastwave. For arbitrary initial conditions there are transients that -- for this combination of ambient power-law index and adiabatic index -- decay with time as $\propto R_0^{-1.32}$ \citep{Coughlin_2022}. Thus the predominant perturbations to the flow are generated by the gravitational field of the compact object, which grow with time.}. We denote Sedov-Taylor quantities with a subscript-0, which encapsulates the fact that these are the ``unperturbed'' variables, such that the Sedov-Taylor shock position and velocity are, respectively, $R_0(t)$ and $V_0(t) = dR_0/dt$. Energy conservation dictates that the Sedov-Taylor solution satisfies (see Section \ref{sec:estimates})
\begin{equation}
    V_0(t) = V_{\rm 0,i}\left(\frac{R_0(t)}{R_{\rm 0,i}}\right)^{-3/4},
\end{equation}
where $V_{\rm 0,i}$ and $R_{\rm 0,i}$ are the initial values for the unperturbed shock velocity and position. A finite gravitational field will then induce variations in the flow. When the shock velocity is much greater than the freefall speed, these variations will be small corrections, which we expect to scale as the ratio of the freefall speed to the shock speed. We therefore expand the shock position as 
\begin{equation}
\begin{split}
    R(t) &= R_0(t)\left(1+\alpha_1 \epsilon_0+\alpha_2\epsilon_0^2+\ldots\right)
    \\
    &= R_0(t)\sum_{n = 0}^{\infty}\alpha_{\rm n}\epsilon_0(\tau_0)^{n}, \label{shockradius}
    \end{split}
\end{equation}
where the $\alpha_{\rm n}$ are as-yet-undetermined coefficients, and 
\begin{equation}
\begin{split}
    \epsilon_0\left(t\right)&=\frac{1}{V_0}\sqrt{\frac{2GM_{\rm ns}}{R_0}} = \frac{1}{V_{\rm 0,i}}\sqrt{\frac{2GM_{\rm ns}}{R_{\rm 0, i}}}\left(\frac{R_0}{R_{\rm 0, i}}\right)^{1/4}
    \\
    &\equiv \epsilon_{\rm i}e^{\tau_0/4}.\label{epsilon}
    \end{split}
\end{equation}
In the final equality we defined
\begin{equation}
    \tau_0 = \ln\left(\frac{R_0(t)}{R_{\rm 0, i}}\right), \quad \epsilon_{\rm i} = \frac{1}{V_{\rm 0, i}}\sqrt{\frac{2GM_{\rm ns}}{R_{\rm 0, i}}}.
\end{equation}
Taking the time derivative of Equation~\eqref{shockradius} yields the shock velocity:
\begin{equation}
    V\left(t\right)=V_0\left(t\right)\sum_{n = 0}^{\infty} \alpha_{\rm n}\left(1+n/4\right)\epsilon_0(\tau_0)^{n}. \label{shock velocity}
\end{equation}

To accommodate the fact that Equations~\eqref{v bc}-\eqref{p bc} take place at a radius that is evolving in time, we make the following change of variables in the fluid equations:
\begin{align}
    \label{xi expression}
    r &\rightarrow \xi, \quad \xi = \frac{r}{R\left(t\right)}, \\ 
    \label{tau expression}
    t &\rightarrow \tau_0, \quad \tau_0 = \ln\left(\frac{R_0\left(t\right)}{R_{\rm 0, i}}\right).
\end{align}
We assume the shock is at sufficiently large radii that the neutron star surface coincides with the origin, $\xi=0$, and the shock is located by construction at $\xi=1$. Analogously to the shock position, we expand the post-shock fluid velocity, density, and pressure as 
\begin{align}
    \label{param velo}
    v &= V\sum_{n = 0}^{\infty} f_{\rm n}(\xi)\epsilon_0(\tau_0)^{n}, \\     
    \label{param dens}
    \rho &= \rho_{\rm i}{\left(\frac{R}{R_{\rm i}}\right)}^{-3/2}\sum_{n = 0}^{\infty}g_{\rm n}(\xi)\epsilon_0(\tau_0)^{n},  \\ 
    \label{param press}
    p &= \rho_{\rm i}{\left(\frac{R}{R_{\rm i}}\right)}^{-3/2}V^2\sum_{n = 0}^{\infty} h_{\rm n}(\xi)\epsilon_0(\tau_0)^{n}. 
\end{align}
Inserting Equations~\eqref{param velo} -- \eqref{param press} into Equations~\eqref{continuity}-\eqref{energy eq}, we equate terms power-by-power in $\epsilon_0$ to construct a set of relations among the various $f_{\rm n}$, $g_{\rm n}$, and $h_{\rm n}$; we perform a similar expansion in the boundary conditions at the shock, Equations \eqref{v bc} -- \eqref{p bc}. While the complete solution has infinitely many terms in the series expansion, in practice we must truncate the series at a finite upper limit, $N$, and hence we have $3N$ relations after equating term-by-term in the three fluid equations alongside $3N$ boundary conditions at the shock. We also expect the first and second order terms to encode different physical information, for while the first-order terms yield corrections at the shock front owing to the momentum flux of the ambient fluid, the second-order (and higher) terms enter the fluid equations directly through the momentum equation via the gravitational term. In Appendix~\ref{sec:appendix} we include the zeroth (i.e., the Sedov-Taylor solution), first, and second-order equations and boundary conditions for the interested reader, but in general we use computer-algebra software to derive the (lengthy) equations.

The $\alpha_{\rm n}$ are constrained by a fourth boundary condition that is not at the location of the shock. In the scenario where a neutron star is the byproduct of the core-collapse, a reasonable boundary condition is that the fluid velocity go to zero as we approach the origin. We therefore have
\begin{equation}
    f_{\rm n}(\xi = 0) = 0,
\end{equation}
and this fourth boundary condition will only be satisfied for certain $\alpha_{\rm n}$. Solving for the $\alpha_{\rm n}$ values is done numerically when solving the fluid equations and the values can be found in Table~\ref{tab:eigenvalues}. 
\begin{table}
\centering
\begin{tabular}{|c|c|c|c|c|c|c|}
\hline 
$\alpha_1$ & \hspace{-.05in} $\alpha_2$  \hspace{-.05in} & \hspace{-.05in} $\alpha_3$ & \hspace{-.05in}  $\alpha_4$ & \hspace{-.05in}  $\alpha_5$ & \hspace{-.05in}  $\alpha_6$ & \hspace{-.05in}  $\alpha_7$ \\
\hline
-0.543 & \hspace{-.05in} -0.0615 & \hspace{-.05in}  -0.0185 & \hspace{-.05in}  -0.00700 & \hspace{-.05in}  -0.00293 & \hspace{-.05in}  -0.00126 & \hspace{-.05in}  -0.000526 \\
\hline
\end{tabular}
\caption{The eigenvalues $\alpha_{\rm n}$ that satisfy the zero-velocity boundary condition at the neutron star surface.}
\label{tab:eigenvalues}
\end{table}

In the solutions we provide here, we restrict ourselves to the scenario that an object with a well-defined surface is formed after core-collapse. However, we expect that a black hole will eventually form, in which case the zero-velocity boundary condition is no longer valid. We will return to the discussion of eventual black hole formation in Section~\ref{sec:BH formation}.

\subsection{Solutions}
\label{sec: solutions}
Equations~\eqref{zeroth cont}-\eqref{second entropy}, as well as the higher order equations, are linear, ordinary differential equations that 
can be solved numerically to determine the $\rm n^{\rm th}$ order dimensionless velocity, density, and pressure profiles ($f_{\rm n}$, $g_{\rm n}$, and $h_{\rm n}$). We focus on the case where $\gamma = 4/3$, as this should be valid during the initial stages of the supernova when the gas temperatures are high and the electrons are relativistic and/or radiation contributes substantially to the pressure support. 
\begin{figure}
    \includegraphics[width=0.47\textwidth]{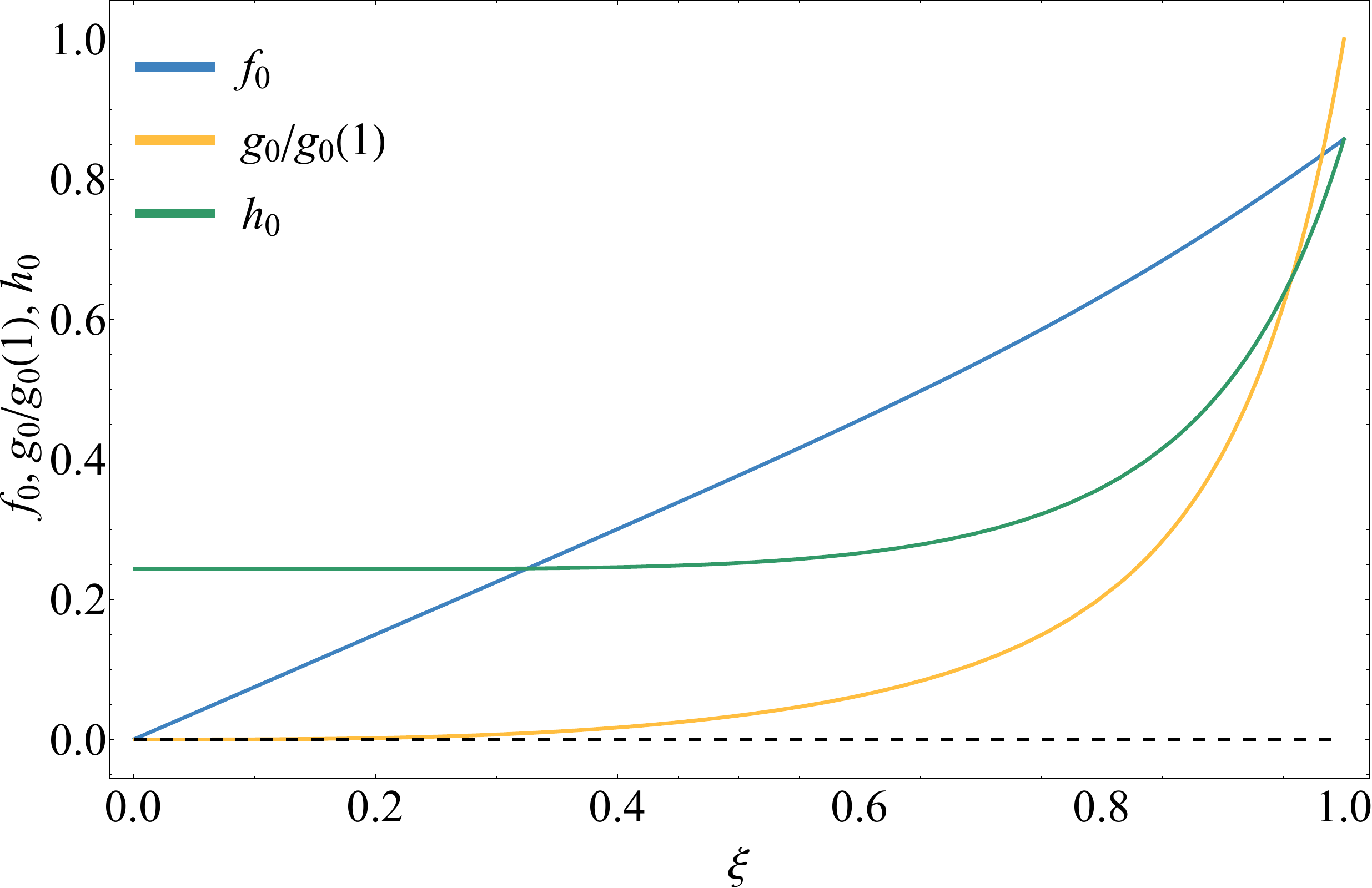}
\caption{The unperturbed, self-similar solution (Sedov-Taylor solution) for an ambient density that falls off as $r^{-3/2}$ and adiabatic index $\gamma=4/3$. The lines correspond to the dimensionless post-shock velocity, density, and pressure as a function of $\xi$, which is the spherical radius $r$ normalized by the shock position. Here we normalize the density by the density at the shock front ($\xi=1$) for clarity.} \label{fig:sedov plot}
\end{figure}
The unperturbed, Sedov-Taylor self-similar solutions -- $f_0$, $g_0$, $h_0$ -- are shown in Figure~\ref{fig:sedov plot}, and the solutions for the corrections, including up to $N = 7$, are shown in Figure~\ref{fig:fgh plots}. Figure~\ref{fig:fgh plots} therefore illustrates the impact of the gravitational field on the post-shock fluid. In comparison to the Sedov-Taylor self-similar velocity, where the fluid interior to the shock is everywhere positive and therefore moves outward with the shock, we see from the left panel of Figure \ref{fig:fgh plots} that the dimensionless velocity profiles are negative everywhere behind the shock, and therefore the motion is toward the origin. At early times, when the Sedov-Taylor solution dominates the flow, this will result in a reduction in the fluid velocity behind the shock. However, at later times the higher order terms will eventually dictate the flow and the fluid velocity interior to the shock will be negative and the gas will fall to smaller radii. The middle panel of Figure~\ref{fig:fgh plots} shows the dimensionless density profiles, from which we see that at early times, when the lower order terms dominate, the majority of the mass is concentrated near the shock at $\xi=1$. However, as $N$ increases, the self-similar density, and therefore the bulk mass, begins to peak in a more central region within the inner flow. These profiles also show that the higher order terms have negative values near the shock front, which shows that, as time increases and the higher order terms begin to dominate, the mass is going to be sapped away in that region and the bulk mass will become more central. We see from the right panel of Figure \ref{fig:fgh plots}, which shows the self-similar pressure profiles that in order to counteract the shift in the location of the bulk mass toward the origin, the pressure profiles increase with $N$, which is necessary to decelerate the flow in the inner region and uphold the zero-velocity boundary condition at $\xi=0$. 

With the solutions for the $f_{\rm n}$, $g_{\rm n}$, and $h_{\rm n}$ up to $N=7$ known, the time varying post-shock fluid velocity, density, and pressure profiles can be solved for using Equations \eqref{param velo}-\eqref{param press}, where the time variation enters through the time-dependence of the shock position, velocity and the smallness parameter $\epsilon_0$. We present solutions to the time-varying post-shock fluid profiles in Section \ref{sec:simulations}, where we compare our analytical prediction to the results of numerical simulations. 
\begin{figure*} 
    \includegraphics[width=0.34\linewidth]{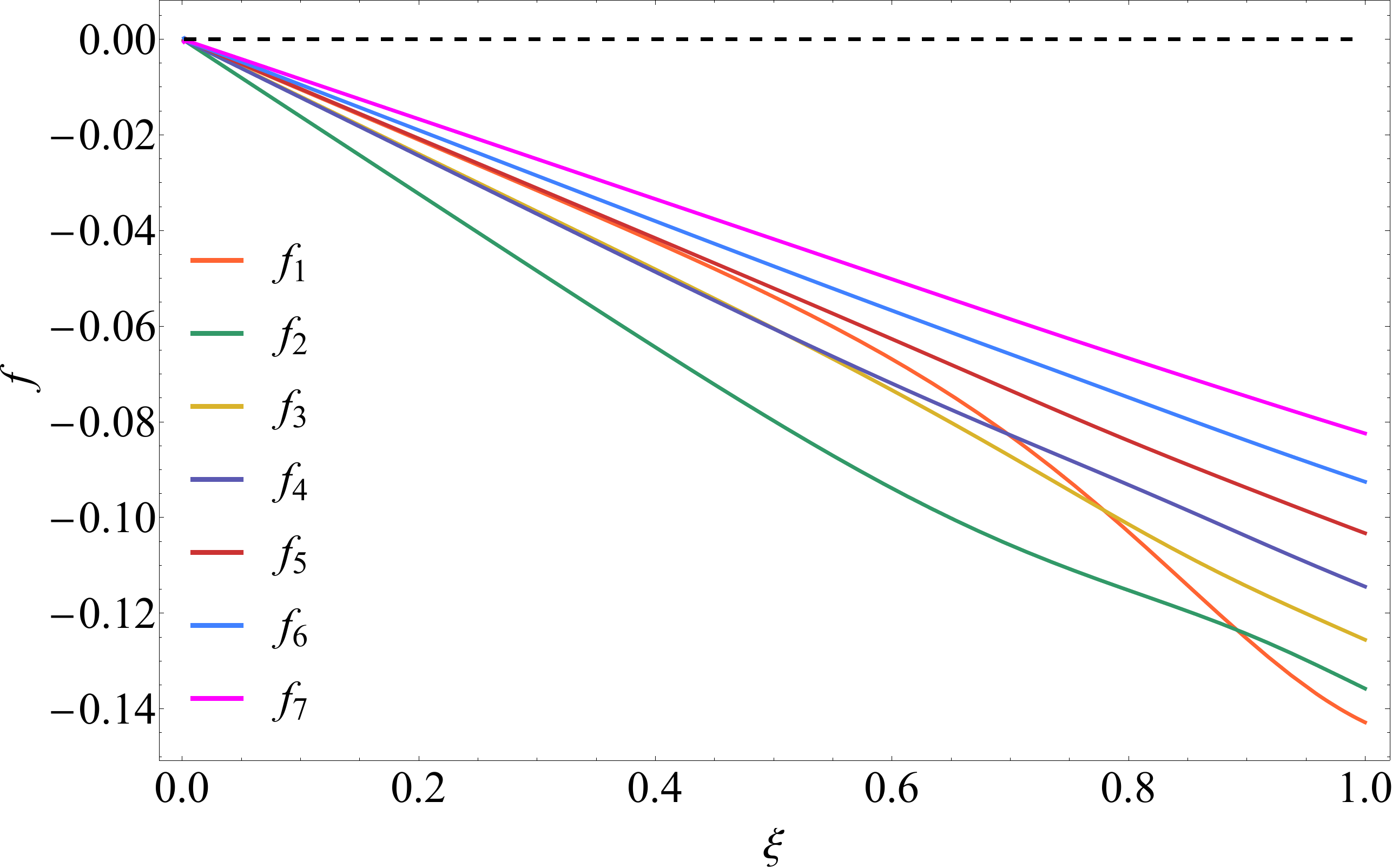}
    \includegraphics[width=0.324\linewidth]{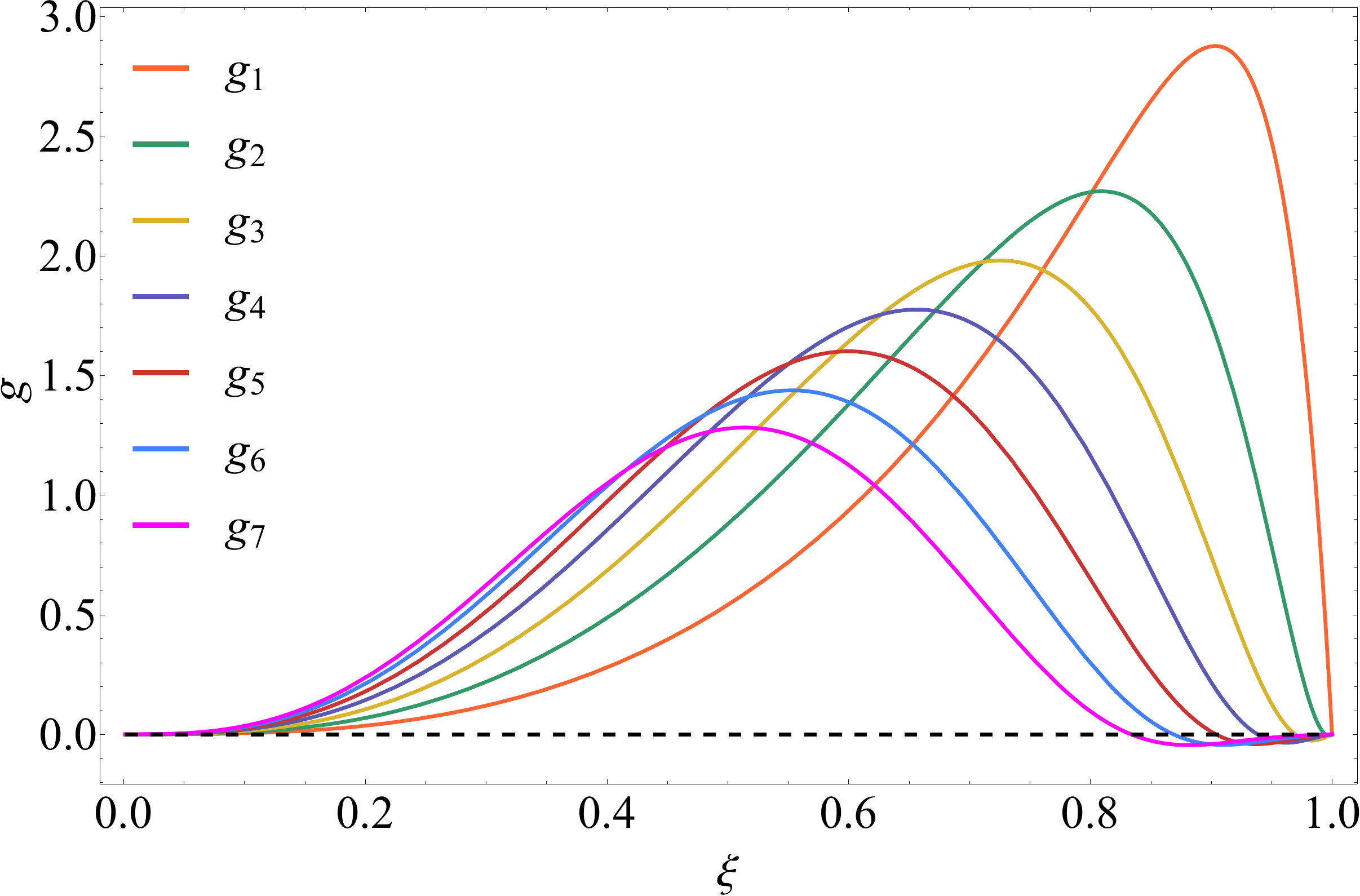}
    \includegraphics[width=0.324\linewidth]{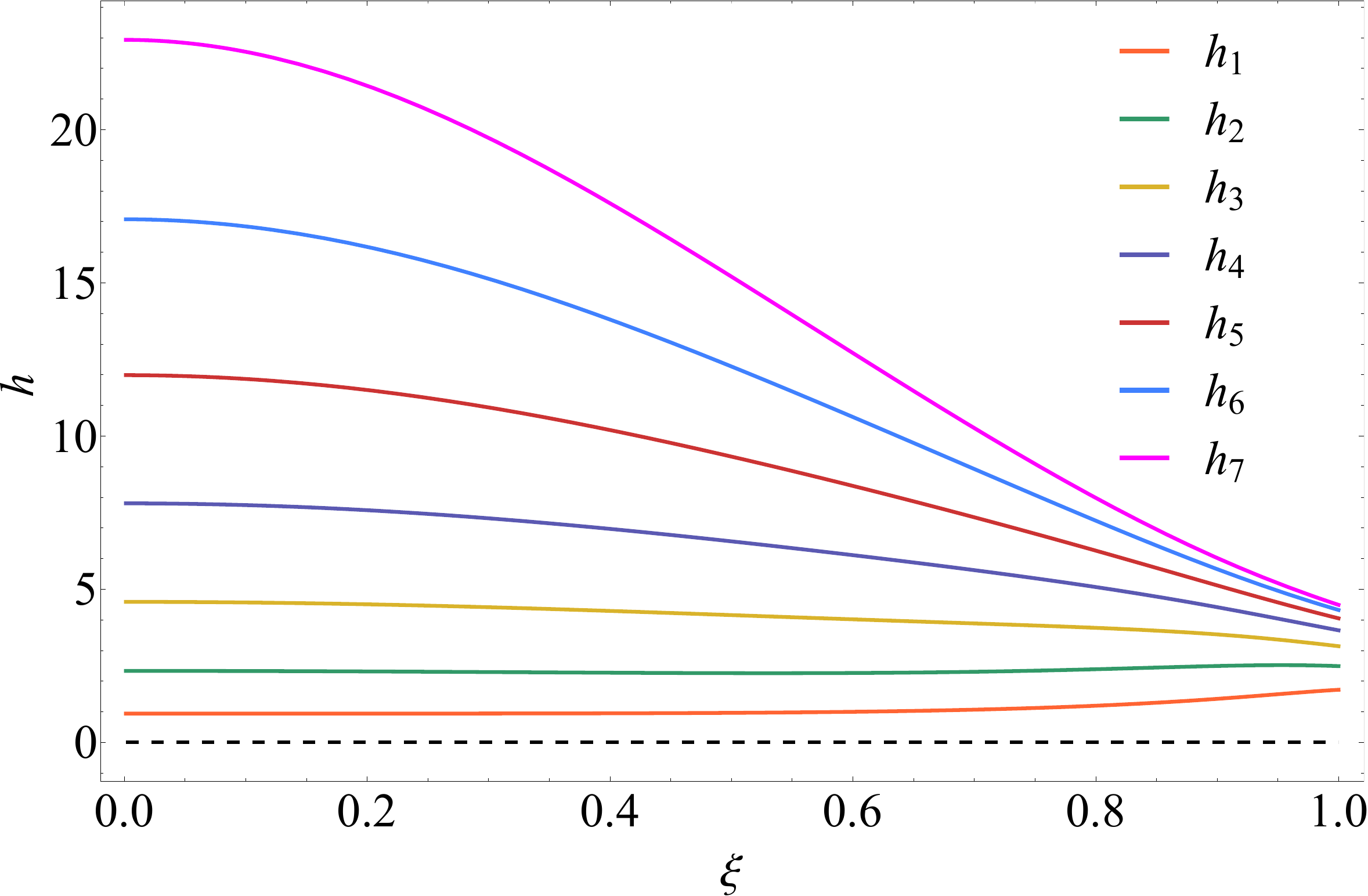}
\caption{The perturbed first through sixth order solutions of $f$, $g$, and $h$ (left, middle, right) for an ambient density that scales as $\rho \propto r^{-3/2}$ with $\gamma=4/3$. The left, middle, and right columns show the dimensionless velocity, density, and pressure of the post-shock fluid as a function of the dimensionless length scale $\xi$. We can see that the velocity profiles become roughly homologous for higher order terms, which therefore motivates the truncation of the series given by Equation~\eqref{param velo} at $N=7$. It can also be seen that higher order terms influence the location of the bulk density. At early times, when the lower order terms dominate, the majority of the post-shock mass is concentrated behind the shock front ($\xi=1$).} \label{fig:fgh plots}
\end{figure*}

\subsection{Stall Radius}
\label{sec:stall}
\begin{figure}
    \includegraphics[width=0.47\textwidth]{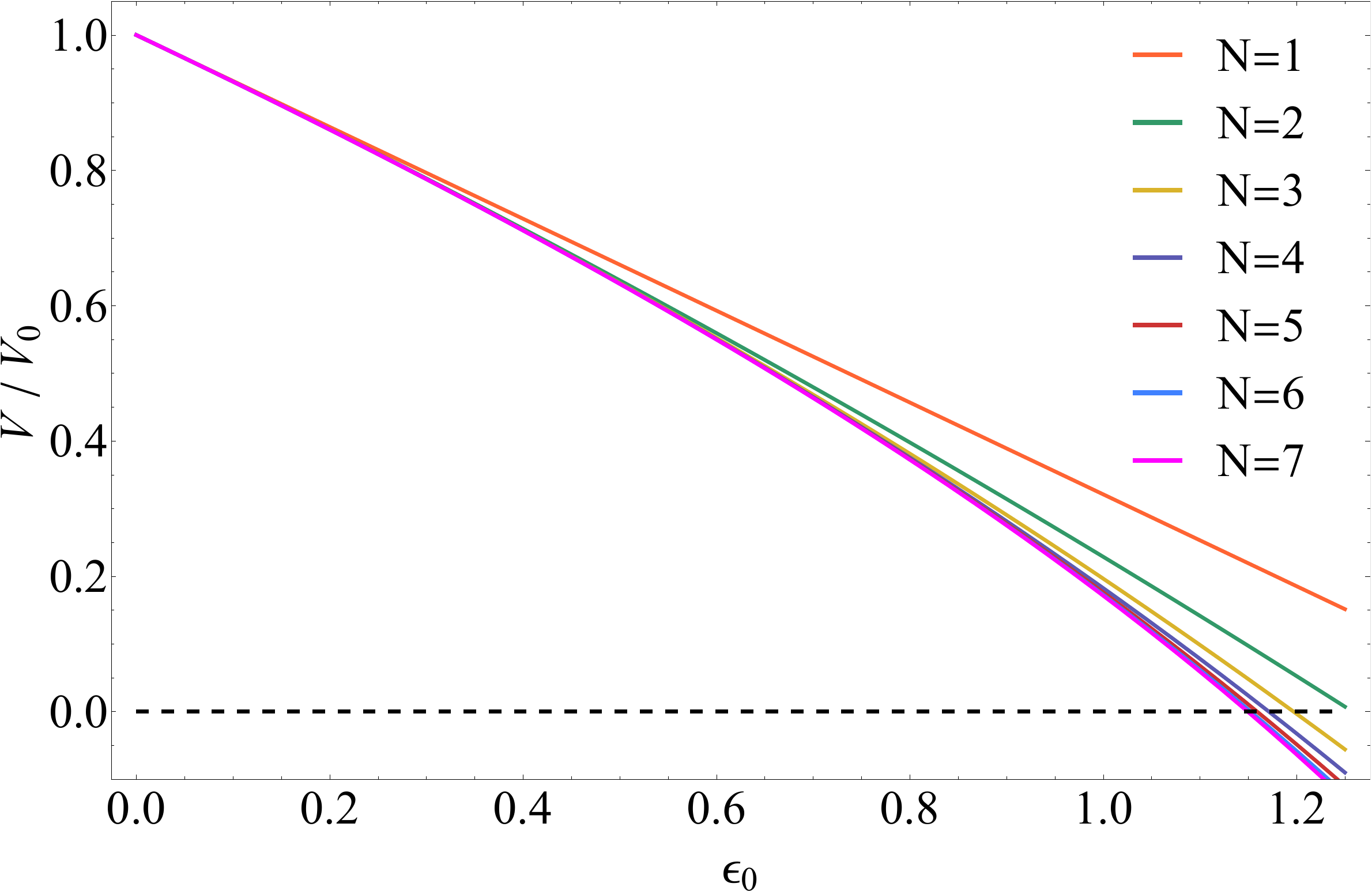}
\caption{Plot of the shock velocity $V$ normalized by the unperturbed shock velocity $V_0$ as a function of the smallness parameter $\epsilon_0$. The shock stalls when $V=0$ and we see that for $N=7$ the shock stalls at $\epsilon_0 \approxeq 1.15$. We can also see that the $\epsilon_0$ value at which the shock stalls is approximately the same for $N=6$ and $N=7$. We therefore truncate the series given by \eqref{shock velocity} at $N=7$ as higher order terms have little impact on the point at which the shock stalls.} \label{fig:VV0 Plot}
\end{figure}
Figure~\ref{fig:VV0 Plot} shows the ratio of the shock velocity to the unperturbed shock velocity obtained from Equation~\eqref{shock velocity} as a function of $\epsilon_0$ and different values of $N$. From this, we can solve for the value of $\epsilon_0$ at which the shock stalls ($V=0$). We see that as the upper limit of the series expansion increases, the zero velocity crossing converges to a specific value of $\epsilon_0$, which we denote $\epsilon_{\rm s}$. Numerically we find
\begin{equation}
    \epsilon_{\rm s} \simeq 1.149.
\end{equation}
We can use Equation \eqref{epsilon} to solve for the unperturbed radius at which the shock stalls, $R_{\rm 0, s}$, being
\begin{equation}
\begin{split}
    R_{\rm 0, s} &= R_{\rm 0, i}\left(\frac{\epsilon_{\rm s}}{\epsilon_{\rm i}}\right)^{4} \\
     \Rightarrow R_{\rm s} &=R_{\rm i}\frac{1+\alpha_1\epsilon_{\rm s}+\alpha_2\epsilon_{\rm s}^2+\ldots}{1+\alpha_1\epsilon_{\rm i}+\alpha_2\epsilon_{\rm i}^2+\ldots}\left(\frac{\epsilon_{\rm s}}{\epsilon_{\rm i}}\right)^{4} \\ 
    &\simeq 0.426 \times R_{\rm i}\frac{\epsilon_{\rm i}^{-4}}{1+\alpha_1\epsilon_{\rm i}+\alpha_2\epsilon_{\rm i}^2+\ldots} \label{Rsofei}
    \end{split}
\end{equation}
Here $R_{\rm s}$ comes from inserting $R_{\rm 0, s}$ into Equation~\eqref{shockradius}. For a given $\epsilon_{\rm i}$, which is approximately the ratio of the freefall speed at the shock radius to the initial shock speed (approximately because it is actually the ratio of the freefall speed at the unperturbed/Sedov-Taylor shock radius to the unperturbed/Sedov-Taylor shock speed), Equation \eqref{Rsofei} yields the position within the medium at which the shock stalls. In agreement with our order-of-magnitude estimates in Section \ref{sec:estimates}, we see that this radius is an extremely sensitive function of $\epsilon_{\rm i}$, the reason being that the ratio of the shock speed to the freefall speed is a very weak function of radius ($\propto R_{\rm 0}^{-1/4}$). Therefore, in changing the value of $\epsilon_{\rm i}$ by a factor of 2, we change the resulting stall radius by a factor of 16, i.e., by more than an order of magnitude. We also see that when $\epsilon_{\rm i} \ll 1$, i.e., when the initial shock speed is much greater than the freefall speed, the polynomial in $\epsilon_{\rm i}$ in the denominator can be set to 1, and Equation \eqref{Rsofei} is identical to Equation \eqref{Rsofeiapp} up to the numerical factor of $0.426$. Therefore, our order-of-magnitude estimates of the stall radius were correct to within a factor of $\sim 2$. 

We can also consider Equation \eqref{Rsofei} as an implicit relation between the necessary $\epsilon_{\rm i}$ for the shock to stall at a given radius $R_{\rm s}$. If $\epsilon_{\rm i}$ is small, which is consistent with our supposition that the shock is initially strong, then the polynomial in the denominator in Equation \eqref{Rsofei} is approximately 1. Making this approximation, the value of $\epsilon_{\rm i}$ in terms of the stall radius is
\begin{equation}
    \epsilon_{\rm i}=0.808 \left(\frac{R_{\rm s}}{R_{\rm i}}\right)^{-1/4}. \label{approx epsiloni}
\end{equation}
Analogously to Equation \eqref{Rsofei}, we see that changing the stall radius -- even by a relatively large factor -- does not dramatically modify the corresponding $\epsilon_{\rm i}$.

\subsection{Blastwave energy and minimum successful explosion energy}
\label{sec:energy}
The total energy behind the blastwave is
 \begin{equation}
     \label{energy integral}
     E=4\pi \int_0^R\left(\frac{1}{2}v^2+\frac{1}{\gamma-1}\frac{p}{\rho}-\frac{GM_{\rm ns}}{r}\right)\rho r^2 dr.
 \end{equation}
We can now insert our expressions for $v$, $\rho$, and $p$ in terms of the dimensionless functions $f$, $g$, and $h$. The result is a sum with (formally) infinitely many terms, the zeroth-order (in $\epsilon_0$) of which is the time-independent contribution from the Sedov self-similar solution, with all latter and time-dependent terms being integrals of the higher-order corrections multiplied by $\epsilon_0(t)^{n}$ with $n \ge 1$. However, in our solution the velocity approaches zero near the origin while the pressure and density remain finite, meaning that the energy flux approaches zero at small radii. Additionally, the ambient medium is in pressureless freefall with $v= -\sqrt{2GM_{\rm ns}/r}$, implying that the energy flux at the shock front is also zero. Therefore, the total energy behind the shock must be a conserved quantity, meaning that the integral multiplying each time-dependent term in the series expansion of the energy -- with the exception of the contribution from the Sedov-Taylor self-similar solution -- must be identically zero. By direct substitution and numerical integration, we have verified that this is the case for the first and second-order terms in the series. 

It thus follows that the energy of the blastwave is
\begin{equation}
\begin{split}
    E = &8\pi\rho_{\rm i}R_{\rm i}^3\frac{GM_{\rm ns}}{\epsilon_{\rm i}^2R_{\rm i}}E_{\star}
    \\
    &\left(1-\frac{\alpha_1}{2}\epsilon_{\rm i}+\frac{1}{8}\left(3\alpha_1^2-4\alpha_2\right)\epsilon_{\rm i}^2+\mathcal{O}\left[\epsilon_{\rm i}^3\right]\right). \label{enint}
    \end{split}
\end{equation}
Here we defined 
\begin{equation}
    E_{\star} \equiv \int_0^{1}\left(\frac{1}{2}f_0^2+\frac{3h_0}{g_0}\right)g_0\xi^2d\xi \simeq 0.601,
\end{equation}
which is a numerical correction introduced by the Sedov-Taylor self-similar solution, and the terms in parentheses arise from the fact that $\epsilon_{\rm i}$ is written in terms of the unperturbed initial shock velocity ($V_{\rm 0i}$) and position ($R_{\rm 0i}$), and hence there is an $\epsilon_{\rm i}$-dependent correction that arises when writing the result in terms of only $R_{\rm i}$. Specifically, the term is $\left(R_{\rm i}/R_{\rm 0i}\right)^{1/2}$, and the second-order-accurate expression in parentheses in Equation \eqref{enint} results from using Equation \eqref{shockradius} to write this ratio in terms of $\epsilon_{\rm i}$. 

Equation \eqref{enint} is useful because, in a realistic explosion, the initial conditions (i.e., when the shock transitions to the strong regime) will not be given by the perturbed Sedov-Taylor solution. However, the energy should still be conserved (until a black hole is formed; see Section \ref{sec:BH formation} below for additional discussion), and the time-dependent terms that account for the initial conditions (i.e., the homogeneous solution to the linearized fluid equations given in Appendix \ref{sec:appendix}) rapidly decay (as $\propto R^{-1.32}$; \citealt{Coughlin_2022}). Therefore, for a known initial energy, Equation \eqref{enint} can be used to determine the corresponding $\epsilon_{\rm i}$ to which that energy corresponds, and hence the time-dependent solution that accounts for the gravitational corrections.

Additionally, from Equation \eqref{approx epsiloni}, we know the minimum $\epsilon_{\rm i}$ necessary to generate a successful explosion by setting the stall radius $R_{\rm s}$ equal to the radius of the progenitor $R_{\star}$. Using this value of $\epsilon_{\rm i}$ and ignoring the higher-order terms in parentheses in Equation \eqref{enint}, which is valid when the shock is initially strong with $\epsilon_{\rm i} \ll 1$, it follows that the minimum energy necessary to drive the shock to the surface of the star is 
\begin{equation}
    E \geq 23.1 \times GM_{\rm ns}\rho_{\rm i}R_{\rm i}^2\left(\frac{R_{\star}}{R_{\rm i}}\right)^{1/2} \label{min energy}. 
\end{equation}
This expression scales identically to Equation~\eqref{energyapprox}, but includes an additional numerical factor of $\sim 23.1$ that results from our more exact treatment. We again take our fiducial progenitor values to be $M_{\rm ns}=1.5$\,\(M_\odot\), $\rho_{\rm i}=10^{-5}~\rm g~cm^{-3}$, ${R_{\rm i}}=10^{12}$\,cm and $R_{\star}=10^{14}$\,cm, and we find that that $E\sim10^{47}$ ergs, which is comparable to the net binding energy of a red supergiant's hydrogen envelope. 

\section{Hydrodynamic Simulations}
\label{sec:simulations}
To test the accuracy of the analytical solutions presented in the previous section and to investigate the time and radius at which the perturbative approach breaks down, we numerically simulate the blastwave evolution with the finite-volume magnetohydrodynamics code {\sc flash} \citep{fryxell00}, version 4.7. We maintain a uniform (i.e., no adaptive mesh), spherical grid throughout each simulation, with a total (fixed) cell number of $2^{16} = 65,536$ and an adiabatic equation of state with $\gamma = 4/3$. The inner boundary is placed at a small but finite radius, our fiducial value being $r_{\rm in} = 0.1$, but we perform tests to assess the sensitivity of the solution to this value. We initialize the fluid variables (density, pressure, velocity) in the simulation with the analytic solutions for a given $\epsilon_{\rm i}$ (with the exception being the test run to analyze the stall location in terms of the energy; see Figure~\ref{fig:setup comparison} below), such that the initial shock position corresponds to $R_{\rm i} = 1$, and the gas is in pressureless freefall for all radii exterior to this radius. The density is set to 1 at a radius just exterior to the shock radius, we set $2GM_{\rm ns} = 1$ (setting the shock radius to 1 and $2GM_{\rm ns} = 1$ is equivalent to normalizing the time by $R_{\rm i}^{3/2}/\sqrt{2GM_{\rm ns}}$), and the pressure floor we use for the ambient medium is $1.0\times 10^{-15}$; this floor value for the pressure can be compared to the initial ram pressure at the shock, being $\sim 1/\epsilon_{\rm i}^2 \gtrsim 10-100$ for the cases we analyze here. 

\begin{figure*} 
    \includegraphics[width=0.33\linewidth]{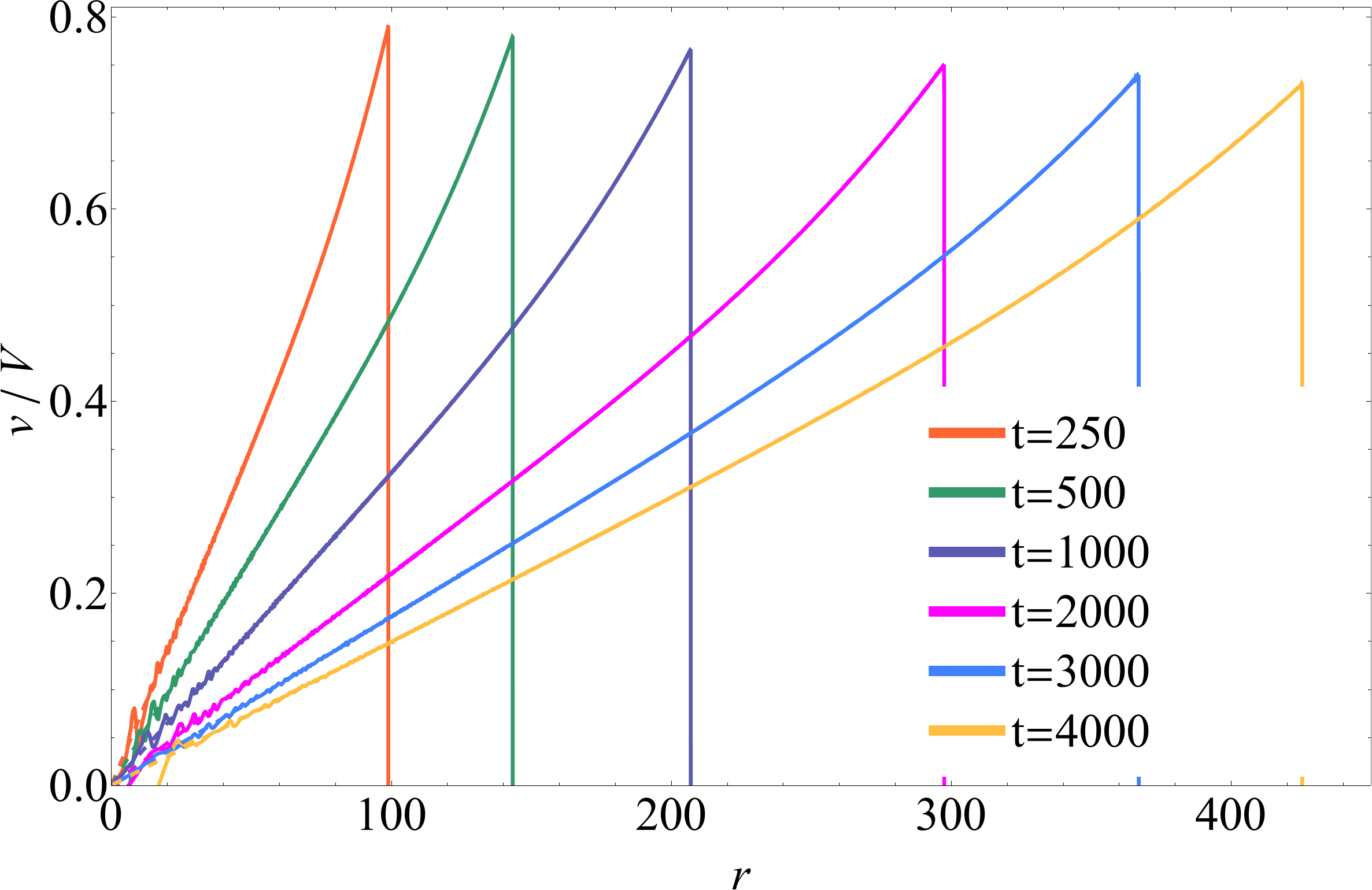}
    \includegraphics[width=0.324\linewidth]{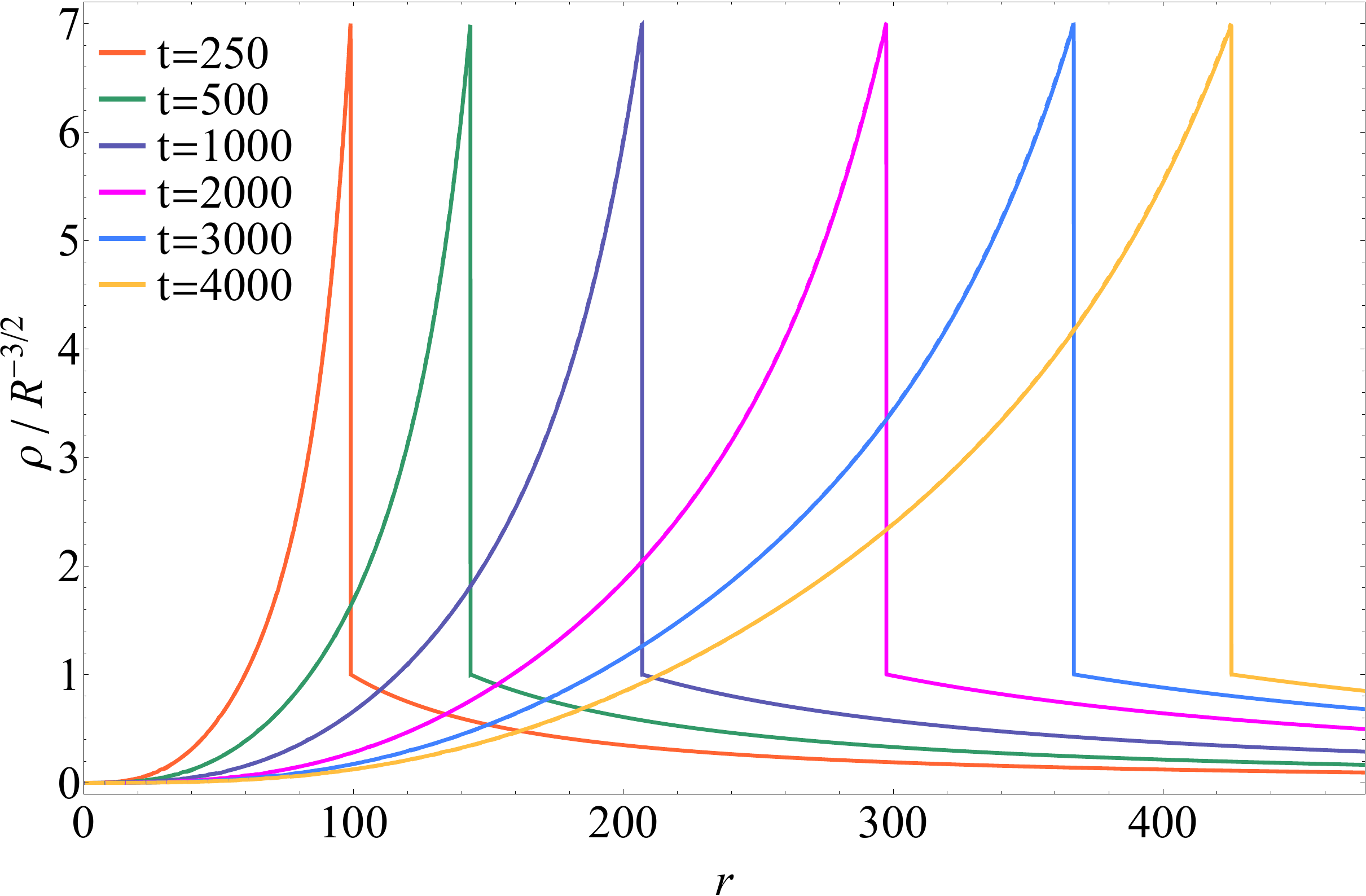}
    \includegraphics[width=0.333\linewidth]{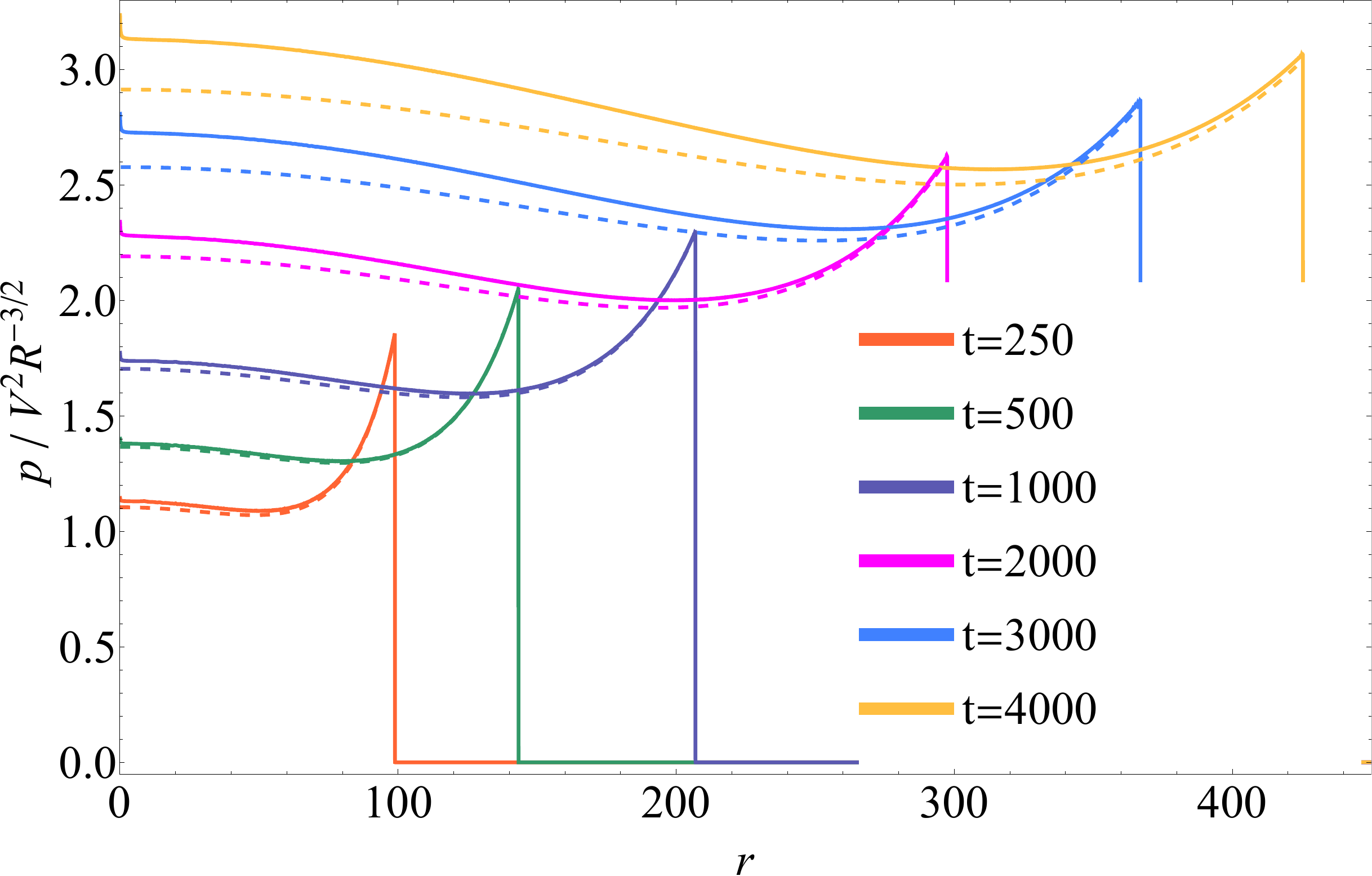}
\caption{Comparison plots between the velocity, density, and pressure profiles (left, middle, right) obtained by {\sc flash} (dashed) and the analytical (solid) prediction normalized by the analytically obtained (through Equations~\eqref{shockradius} \& \eqref{shock velocity}) shock variables for $\epsilon_i=0.1$. The analytically predicted dimensionless velocity profile is nearly indistinguishable from the numerical results, besides the fluctuations due to numerical noise at small radii. It can be seen that at later times the analytically predicted dimensionless pressure is lagging behind the numerical results in the inner region which is due to the neglect of higher order ($N>7$) terms in the series expansion.} \label{fig:op1 comoparison plots}
\end{figure*}
Figure~\ref{fig:op1 comoparison plots} shows the velocity (left), density (middle), and pressure (right), normalized by the shock velocity, ambient density at the shock, and ambient ram pressure at the shock, respectively, for the times in the legend and for $\epsilon_{\rm i} = 0.1$. The black curves show the results of the numerical simulations, while the colored curves give the analytical predictions, including terms up to 7th order in $\epsilon_0(t)$. Note that the shock position and velocity (i.e., the values of $V$ and $R$ that are used to normalize the fluid variables) are determined from the analytical solution, i.e., these are determined self-consistently and there are no free parameters here. It is clear that the numerically obtained velocity and density match the analytical predictions extremely well. At early times the pressure from the numerical simulations is also indistinguishable from the analytical prediction, but at later times the numerical value near the origin is somewhat larger than the analytical prediction. This discrepancy arises from the fact that higher-order terms in the series expansion are needed to accurately recover the pressure near the origin.
\begin{figure*} 
    \includegraphics[width=0.49\linewidth]{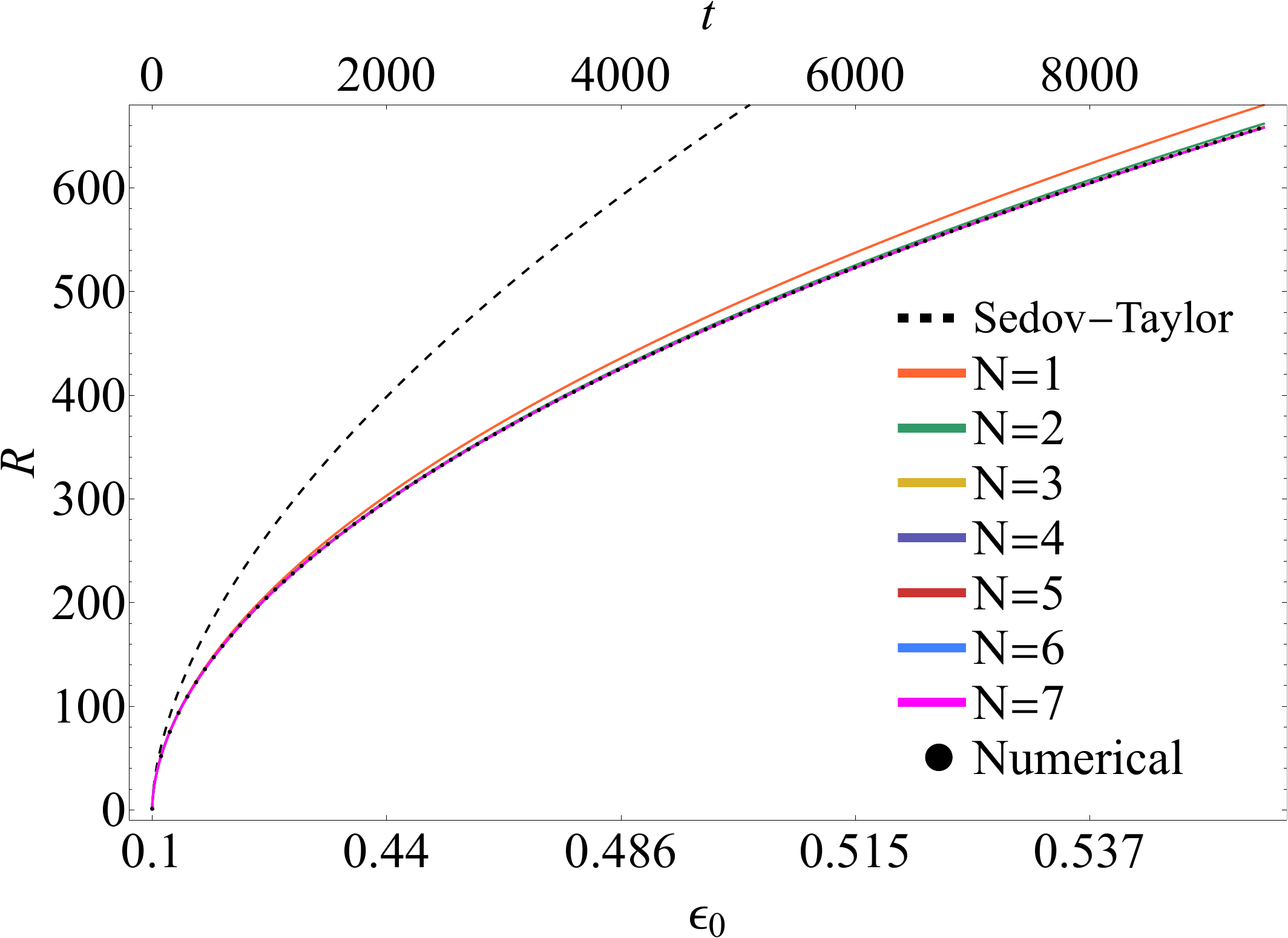}
    \includegraphics[width=0.505\linewidth]{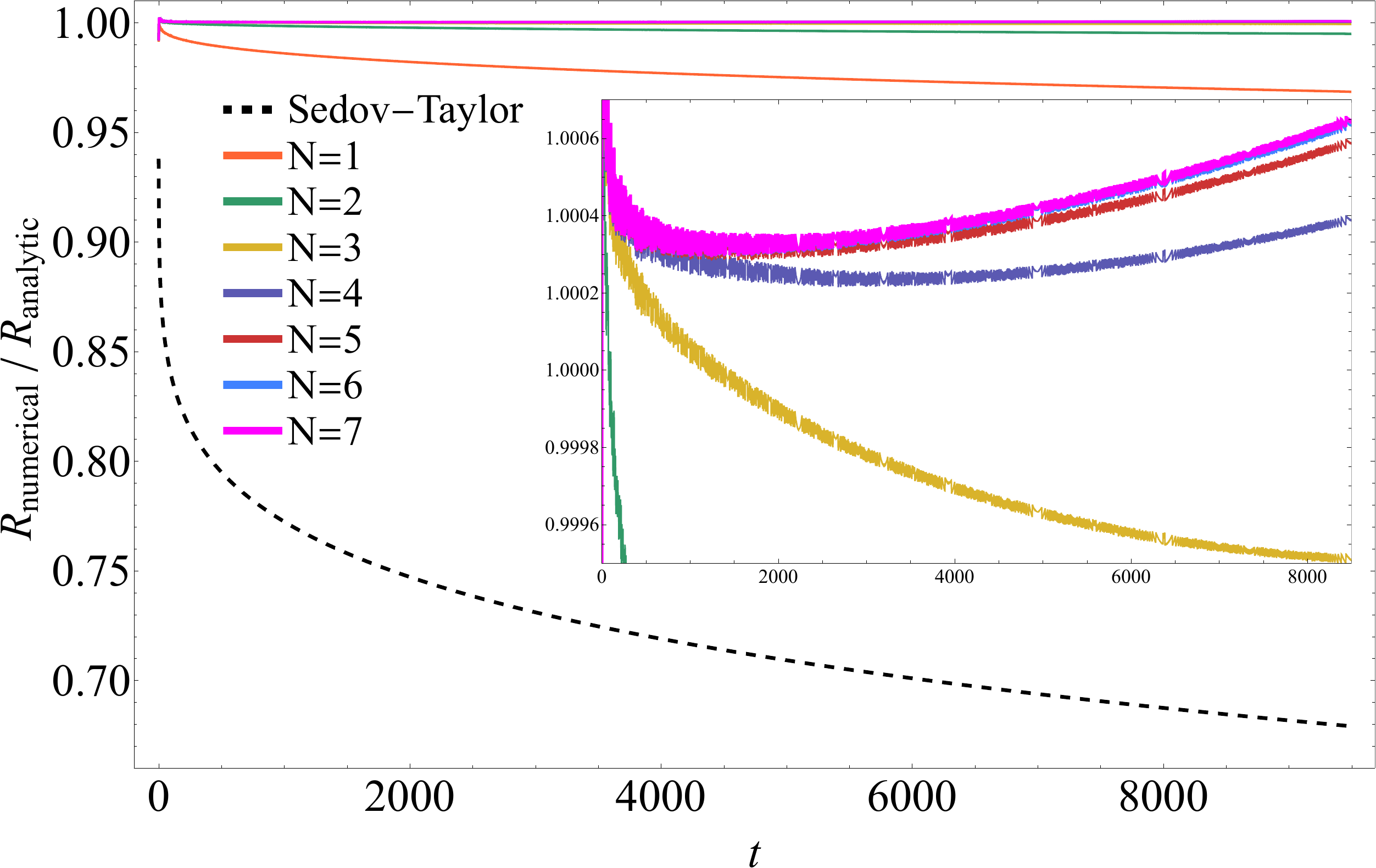}
\caption{Left: Comparison between the analytical prediction for the shock position given by Equation~\eqref{shockradius} and the numerically obtained shock position (black dots) for different values of $N$ as a function of time for $\epsilon_{\rm i}=0.1$. The dashed line shows the unperturbed shock position. Right: The ratio of the numerical shock position to the analytical shock position for different values of $N$. The inset plot is a zoomed in view, which shows that the ratio is very close to unity.} \label{fig:op1 R plots}
\end{figure*}
The left panel of Figure~\ref{fig:op1 R plots} shows the shock position $R(t)$ as a function of $\epsilon_0$ on the bottom-horizontal axis and time on the top-horizontal axis for the analytical prediction (and different values of $N$ as given by Equation~\eqref{shockradius}) and the numerical simulation. The analytical and numerical results are clearly in extremely good agreement. The plot on the right shows the ratio of the two values at different times, which is equal to $1\pm 10^{-3}$ at all times for the analytic solutions that have $N \ge 3$. On the contrary, the Sedov-Taylor solution quickly deviates from the numerically obtained (and higher-order analytical) values, and hence the inclusion of the gravitational terms is necessary for accurately constraining the position of the shock with time.

This figure shows that the analytical approach very accurately predicts the position of the shock, even as $\epsilon_0$ approaches values that are not dissimilar from unity. However, $\epsilon_0$ increases as an extremely shallow power low in time -- since $\epsilon_{0} \propto R_0^{1/4}$ and the Sedov-Taylor shock position satisfies $R_0 \propto t^{2/(5-n)} \propto t^{4/7}$, we have $\epsilon_0 \propto t^{1/7}$. Therefore, while $\epsilon_0$ increases from 0.1 to 0.5 in $\sim 10^{4}$ in time, it would require running the simulation to $\sim 10^7$ in time before the shock is predicted to stall. At our fiducial resolution and without adaptive mesh, running the simulation to this late of a time is infeasible, even in one dimension. 

\begin{figure*} 
    \includegraphics[width=0.495\linewidth]{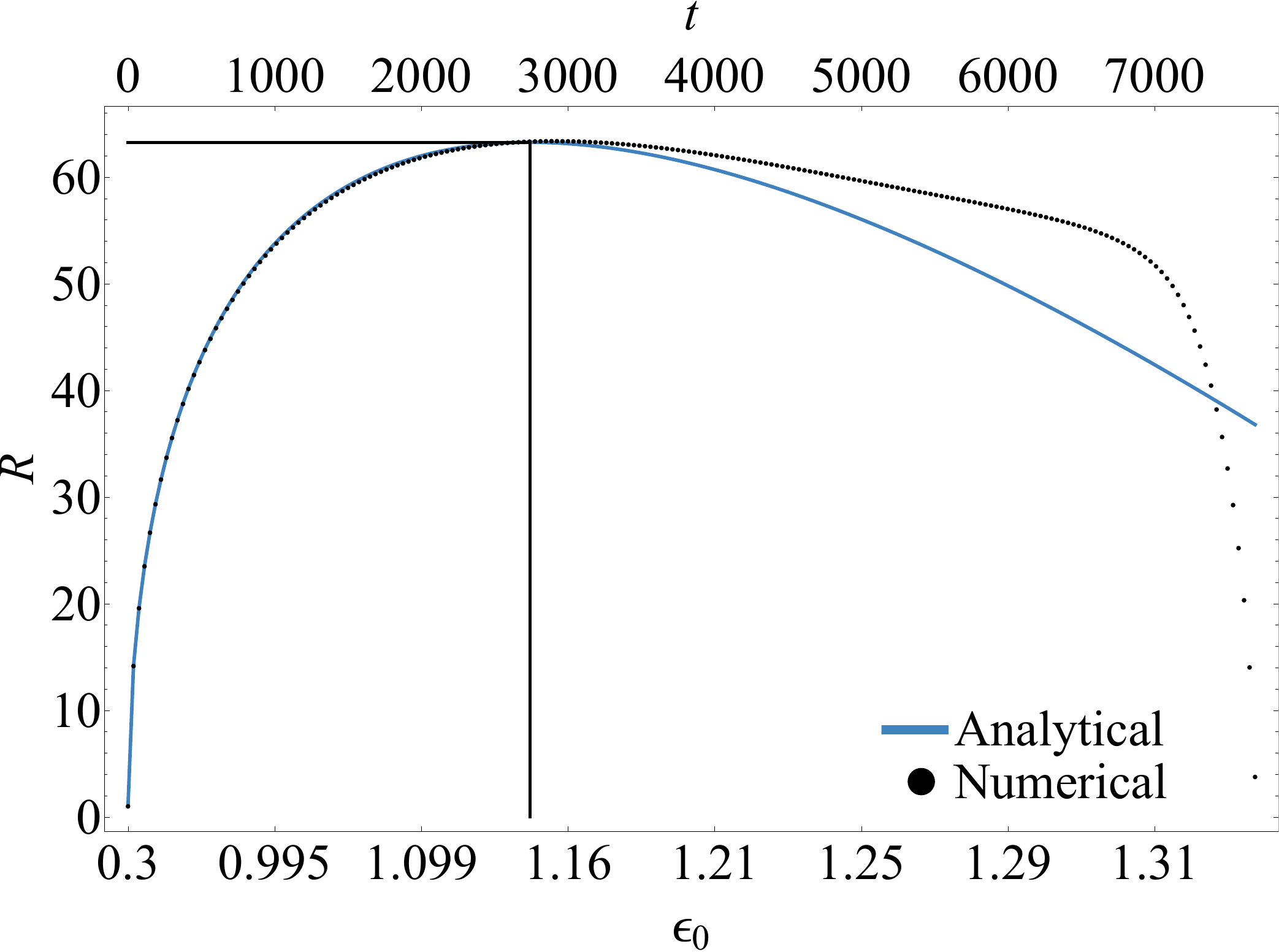}
    \includegraphics[width=0.495\linewidth]{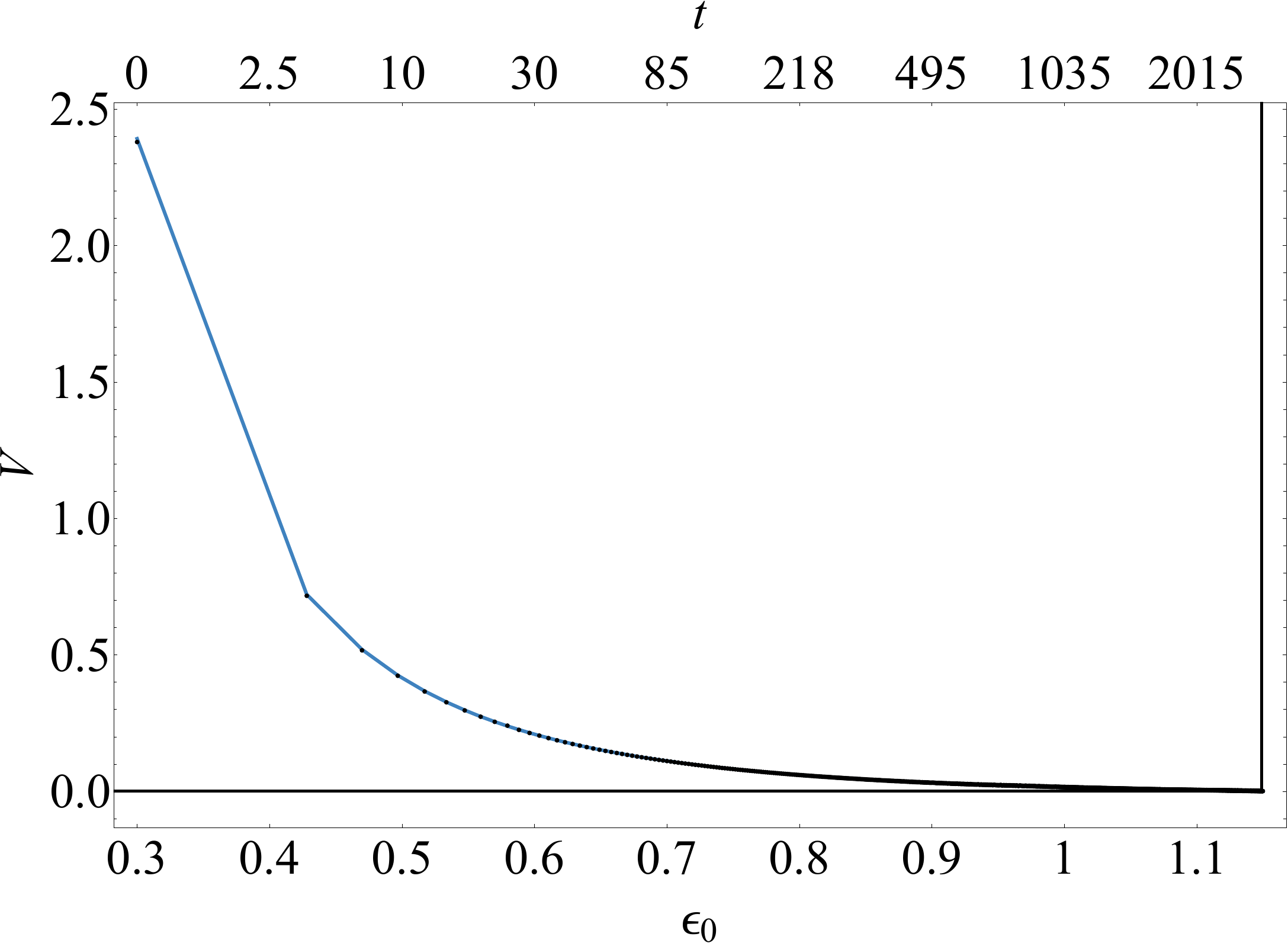}
\caption{Left: Plot of the numerically obtained shock position as a function of time obtained from the hydrodynamics code {\sc flash} (dotted) compared to the analytical prediction of the shock position with an initial ratio of the freefall speed to the shock velocity of $\epsilon_{\rm i}=0.3$. The horizontal and vertical black lines show the analytically predicted radius at which the shock stalls, $R_{\rm s}$, and the corresponding smallness parameter value, $\epsilon_{\rm s}$. Right: The same thing but the shock velocity as a function of  $\epsilon_0$ (and time).} \label{fig:0p3 plots}
\end{figure*}
However, Figures \ref{fig:op1 comoparison plots} and \ref{fig:op1 R plots} also show that the analytical prediction is accurate even when $\epsilon_{\rm 0}$ is larger by a factor of a few than the initial value of $0.1$. Therefore, we can numerically investigate the transition of the shock to its stalled state by artificially increasing the initial value of $\epsilon_0$, as the shock will stall substantially sooner and in a time that is numerically feasible if $\epsilon_{\rm i}$ is large enough (see Equation \ref{Rsofei}). Figure~\ref{fig:0p3 plots} shows the shock position (left) and velocity (right) when $\epsilon_{\rm i} = 0.3$. 
The horizontal and vertical lines show the analytically predicted values for which the shock stalls, being $R_{\rm s} \simeq 63.3$ and $\epsilon_0 = \epsilon_{\rm s} \simeq 1.149$ ($t \simeq 2740$). We see that the analytical prediction very accurately traces the numerical solution to times just after the shock stalls, but thereafter there are noticeable deviations that cannot be captured via the series approach. In other words, the radius of convergence of the series coincides with $\epsilon_{0} \simeq \epsilon_{\rm s}$, and thereafter the numerical simulations are necessary to determine the fate of the shock. 

\begin{figure} 
    \includegraphics[width=0.47\textwidth]{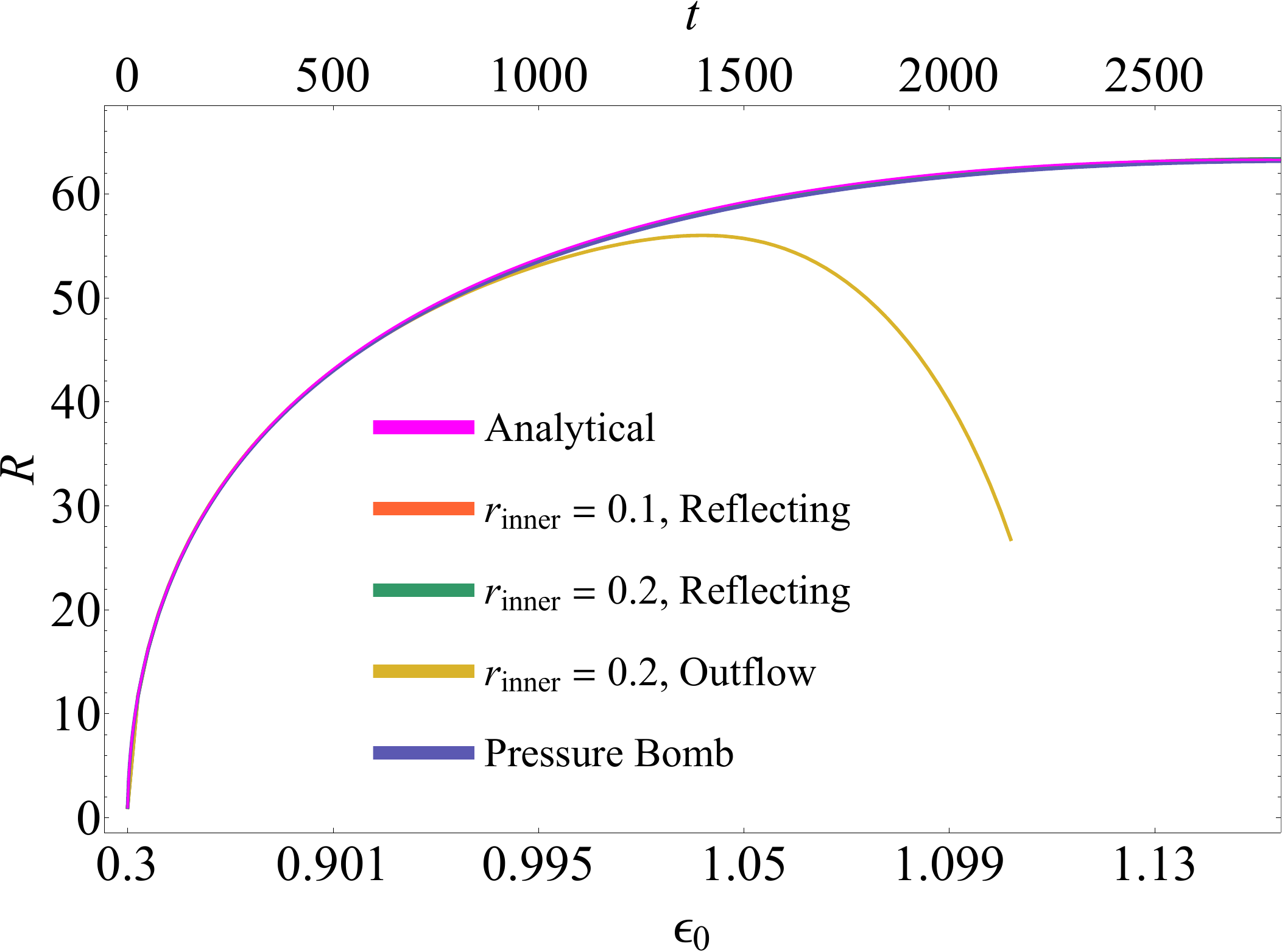}
\caption{Comparison of different simulation setups for $\epsilon_{\rm i}=0.3$. This shows that the simulation results are insensitive to the inner radius, as well as to the initial conditions. The "Pressure Bomb" simulation comes from setting the initial pressure interior to the initial shock position to a uniform value, the velocity to zero, and the density to 0.01. The uniform pressure, $p_{\rm i}$, can be calculated through Equation~\eqref{enint}, where the energy of the explosion is given by the integral of the specific energy, $4\pi p_{\rm i}$. This shows that what dictates the behavior of these explosions is the total energy.} \label{fig:setup comparison}
\end{figure}
Figure~\ref{fig:setup comparison} illustrates the impact of both the boundary and initial conditions on the numerically obtained shock position when $\epsilon_{\rm i}=0.3$. The solution with $r_{\rm inner} = 0.1$ with a reflecting boundary condition (orange curve) is our fiducial setup, the green curve uses a larger inner boundary by a factor of 2, and the yellow curve maintains the same inner radius as the fiducial case but uses an outflow (i.e., zero gradient) boundary condition instead of a reflecting boundary condition. All three of these cases -- the orange, green, and yellow curves -- use the analytical prediction as the initial condition. The blue curve uses a ``pressure bomb'' for the initial conditions, such that the velocity everywhere interior to $R = 1$ is zero, the density interior to $R = 1$ is $0.01$, and the pressure is set to a constant $p_{\rm i}$. The value of $p_{\rm i}$ is set by equating the energy in Equation~\eqref{enint} with $\epsilon_{\rm i} = 0.3$ to the integral of the specific energy behind the shock, which is just $4\pi p_{\rm i}$ for the pressure bomb. This figure shows that changing the value of the inner radius, but maintaining a reflecting inner boundary condition, results in effectively no change in the propagation of the shock. Perhaps surprisingly (given how different the initial conditions are to the perturbed Sedov solution), the simulation with the pressure bomb as the initial condition also yields almost no discernible difference in the shock propagation as compared to the fiducial setup. Using an outflow boundary condition in the interior does have an impact on the solution, and in particular the shock stalls at a noticeably earlier time and smaller radius. These results suggest that the energy of the explosion is what predominantly influences the shock propagation, not the initial conditions. In the case of the outflow boundary condition, energy is allowed to leave the domain -- and this effect becomes most significant as the shock decelerates and more mass is concentrated near the origin -- and causes the shock to decelerate more rapidly.

\subsection{Settling Solution}
The velocity, density, and pressure of the fluid interior to an indefinitely-stalled shock (i.e., one for which the velocity is zero and time-steady) can be described by the self-similar solutions of \citet{Lidov57, chevalier89, Blondin03} (see also \citealt{kundu22} for the general relativistic extension). In these solutions, the velocity of the fluid goes to zero as it approaches the origin and ``settles'' onto the surface of the neutron star, and it is interesting to compare the analytical settling solution to the numerical results once the shock stalls. Following the methods outlined by \citet{Blondin03}, we write the fluid velocity, density, and pressure as 
\begin{align}
    v&=\sqrt{\frac{2GM_{\rm ns}}{R_{\rm s}}} f\left(\xi\right), \label{v settling} \\
    \rho&={R_{\rm s}}^{-3/2} g\left(\xi\right), \label{rho settling} \\
    p&=\frac{2GM_{\rm ns}}{R_{\rm s}}{R_{\rm s}}^{-3/2} h\left(\xi\right), \label{p settling}
\end{align}
where $\xi=r/R_{\rm s}$ is, as above, the dimensionless spherical radius normalized by the shock position $R_{\rm s}$, but now $R_{\rm s}$ is the stalled shock location and is therefore independent of time. We normalize our solution by letting $\dot{M}=4\pi$ and $2Gm=1$ and, from the jump conditions \eqref{v bc} -- \eqref{p bc} with $V \equiv 0$, the self-similar fluid velocity, density, and pressure just behind the shock front are
\begin{align}
    f\left(1\right)&=-\frac{\gamma-1}{\gamma+1}, \label{fsettling bc} \\
    g\left(1\right)&=\frac{\gamma+1}{\gamma-1}, \label{gsettling bc} \\
    h\left(1\right)&=\frac{2}{\gamma+1}. \label{hsettling bc}
\end{align}
The time-steady, self-similar continuity, radial momentum, and entropy equations that describe the fluid interior to the stalled shock are then determined by inserting Equations \eqref{v settling}-\eqref{p settling} into Equations \eqref{continuity}-\eqref{energy eq} with all time derivative terms equal to zero. Doing so, and performing some algebraic manipulations gives
\begin{align}
\label{settling continuity}
    \frac{\partial}{\partial{\xi}}[\xi^2gf]&=0, \\
\label{settling momentum}
    f\frac{\partial{f}}{\partial{\xi}}+\frac{1}{g}\frac{\partial{h}}{\partial{\xi}}&=-\frac{1}{2\xi^2}, \\
\label{settling entropy}
    \frac{\partial}{\partial{\xi}}\ln\left(\frac{h}{g^{\gamma}}\right)&=0. 
\end{align}
Equations \eqref{settling continuity}-\eqref{settling entropy} can therefore be solved numerically by integrating inward from the shock front at which point the dimensionless fluid variables satisfy the boundary conditions given by Equations \eqref{fsettling bc}-\eqref{hsettling bc}.  
\begin{figure*} 
    \includegraphics[width=0.33\linewidth]{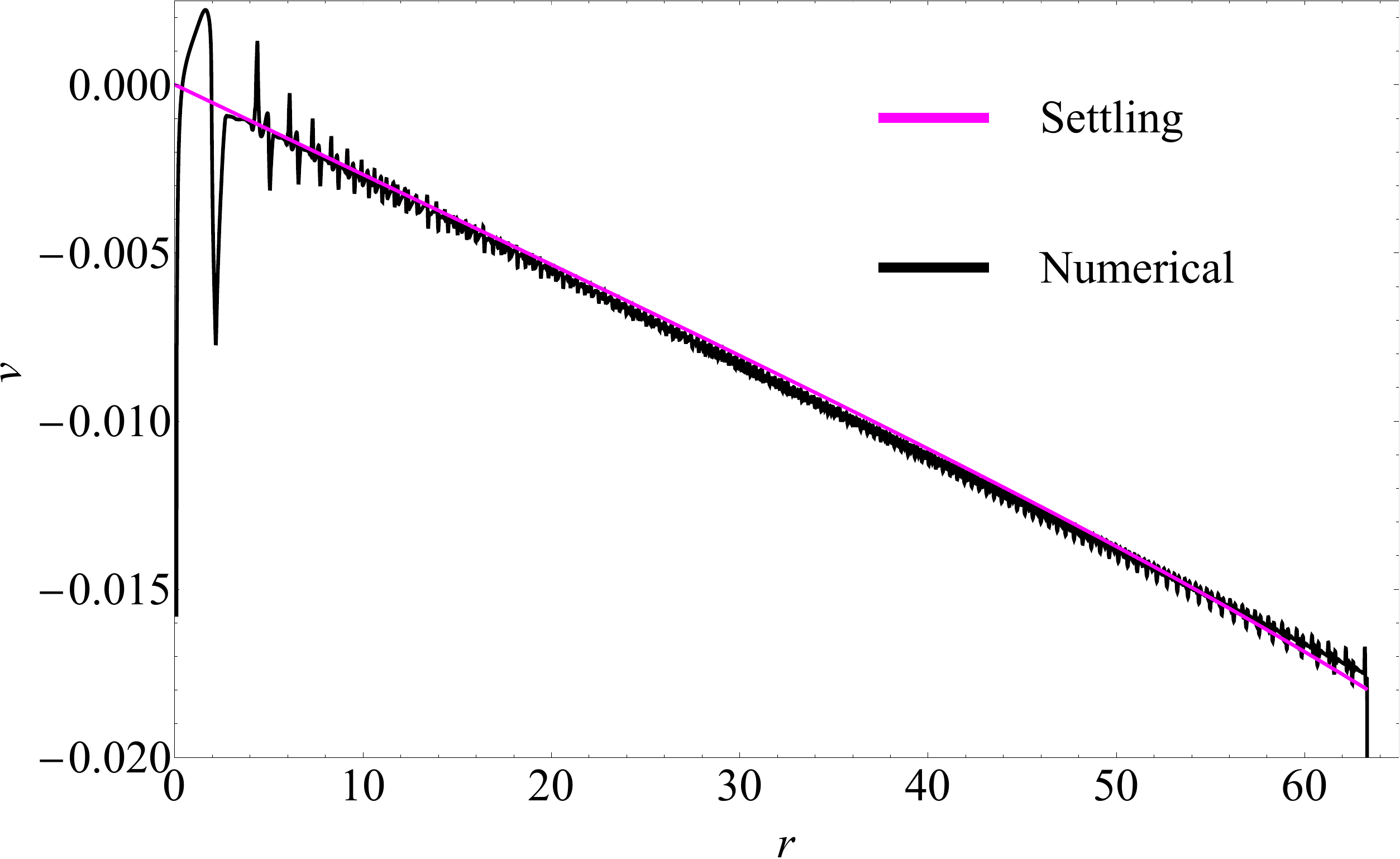}
    \includegraphics[width=0.324\linewidth]{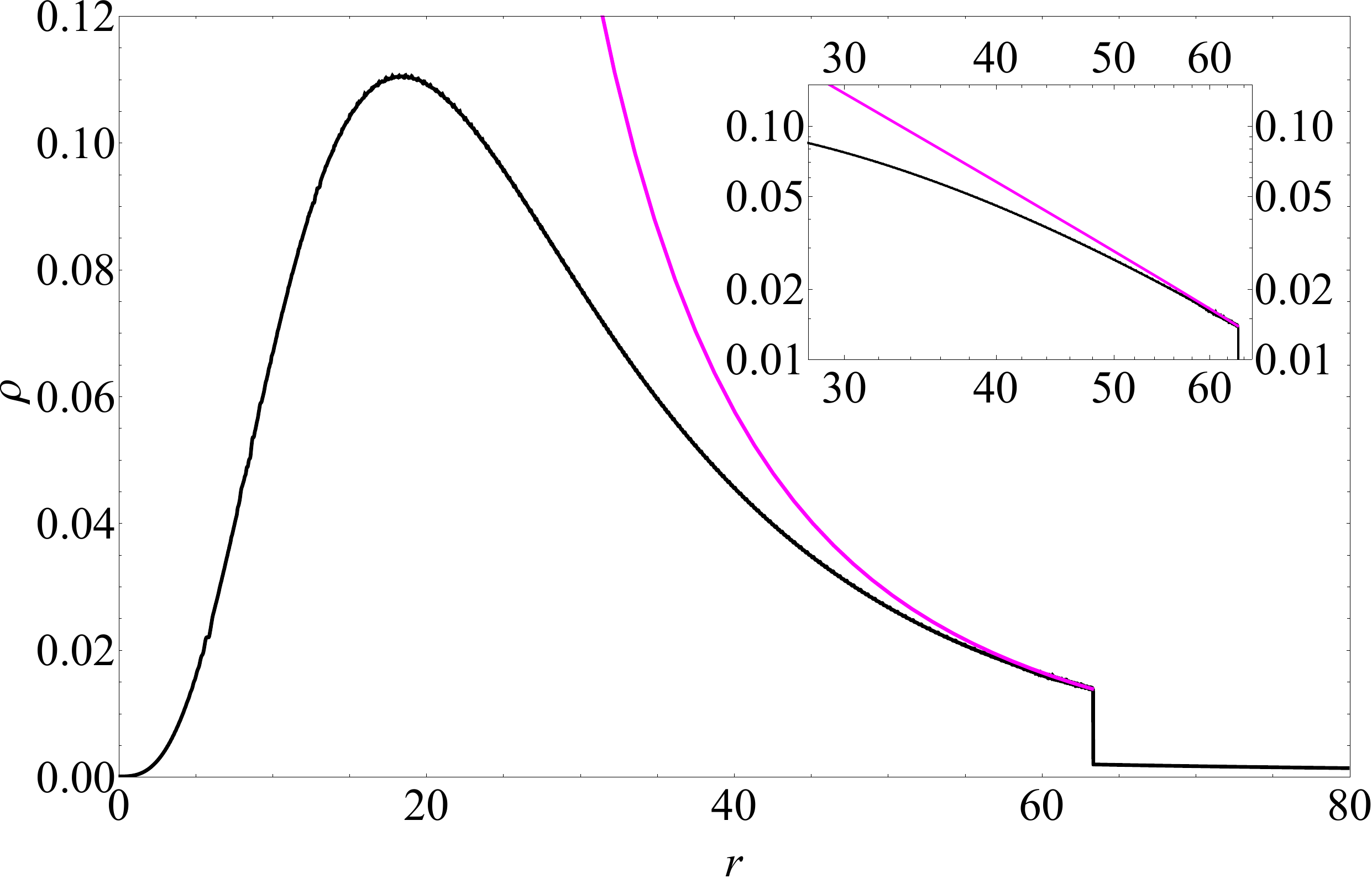}
    \includegraphics[width=0.324\linewidth]{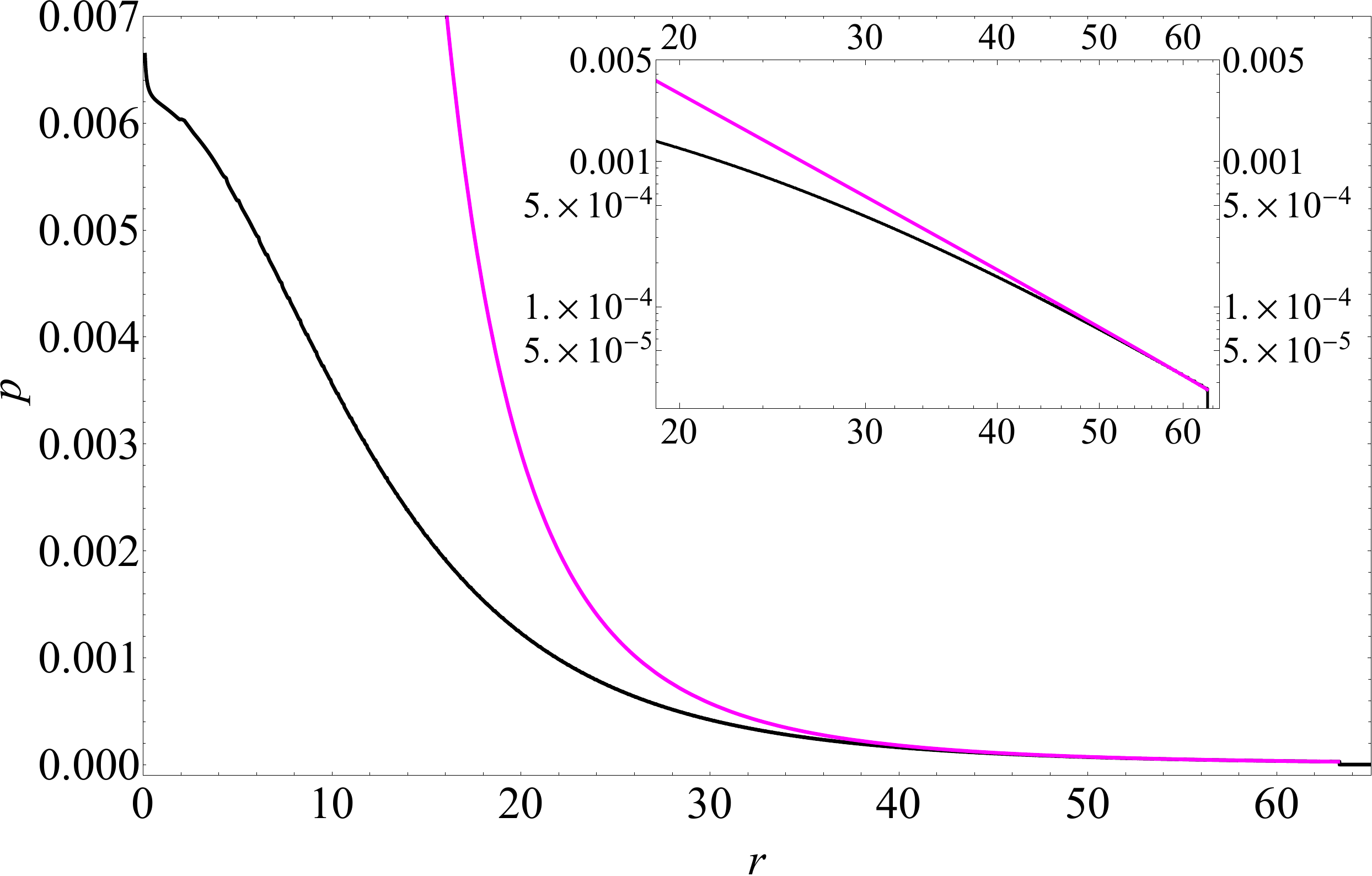}
\caption{Comparison plots between the settling solutions and the numerical profiles given by {\sc flash} for $\epsilon_{\rm i}=0.3$ at $t=2750$, which is just after the time that the shock is predicted to stall. Here we see that the numerical velocity profile very nearly matches onto the analytical settling solution, while the density and pressure only match onto the settling solution for regions near the (stalled) shock front. The inset plots for the density (middle) and pressure (right) show the density and pressure on a log-log scale, which better shows how quickly the numerical solution diverges from the settling solution.} \label{fig:settling plots}
\end{figure*}

For the case of $\gamma=4/3$, the fluid variables satisfy the following scaling
\begin{align}
    v\left(r\right)&\propto r, \\
    \rho\left(r\right)&\propto r^{-3}, \\
    p\left(r\right)&\propto r^{-4},
\end{align}
\citep{chevalier89, Blondin03} and in Figure~\ref{fig:settling plots} we show the comparison between the analytical settling solutions and the numerical solution given by {\sc flash} at $t=2750$ (i.e., when the shock stalls) where the analytical solutions are given by Equations \eqref{v settling}-\eqref{p settling} with $R_{\rm s}=63.3$ and $2GM=1$.
Although our analytical model cannot account for the late time behavior ($t\gtrsim t_{\rm stall}$) of the shock -- as is evident in Figure~\ref{fig:0p3 plots} -- we see from the numerical simulation that the shock attempts to stall indefinitely, but eventually retreats to smaller radii. In order to maintain the zero-velocity boundary condition in the interior, the pressure would need to continuously rise until eventually an infinite pressure gradient 
is necessary to decelerate the flow. This can most clearly be seen in Figure~\ref{fig:settling plots}, where we can see that the fluid velocity, density, and pressure \textit{attempt} to match onto the settling solution, but the pressure gradient is not sufficient enough to withstand the accumulation of gas in the interior. Therefore, it is around the time at which the shock stalls that we predict a black hole will form and the velocity eventually goes to $\sim$ freefall in the interior.

\section{Discussion and Observational Implications}
\label{sec:discussion}
Here we discuss some observational implications of our work, particularly in regard to the newly observed class of Fast X-ray Transients (FXTs), as well as caveats (i.e., approximations and assumptions). 

\subsection{Black hole formation}
\label{sec:BH formation}
In the event of a core-collapse supernova, the formation of a black hole in the central region is always preceded by the neutron star phase. However, a black hole forms once the neutron star is pushed over the Tolman–Oppenheimer–Volkoff (TOV) limit. From both our analytical and numerical solutions (see Figure~\ref{fig:settling plots}), we see that the density is very low near the origin until the shock stalls. This therefore suggests the neutron star will not reach the TOV limit (and a black hole will not form) until very near (and slightly after) the time at which the shock stalls. 

Once a black hole forms, a sound wave propagates into the fluid at a time-dependent position $R_{\rm s}(t)$ that informs the gas of the presence of the black hole at the origin. Therefore, one could model this scenario by using the same series-expansion approach but enforcing that the fluid velocity equal zero at the time-dependent sonic radius, i.e., all the fluid inside of this radius should be in $\sim$ freefall onto the black hole\footnote{Our solution is unable to account for the transition to supersonic freefall onto the black hole due to the nonlinear terms that give rise to this behavior. Said another way, our series solution cannot self-consistently treat the inner regions of the flow for which the gravitational terms dominate the dynamics.} . The dimensionless sonic radius $\xi_{\rm s} = R_{\rm s}/R$ can be determined from an analysis of the characteristics of the linearized fluid equations about the Sedov-Taylor blastwave, and is implicitly given by \citep{Coughlin_2022} 
\begin{equation}\tau_0=\int_{0}^{\xi_{s}\left(\tau_0\right)}\frac{d\xi}{f_0-\xi+\sqrt{4 h_0/(3g_0)}}. \label{sc time}
\end{equation}
Here $f_0$, $h_0$, and $g_0$ are the self-similar solutions for the velocity, pressure, and density of the Sedov-Taylor solution, and $\tau_0 = R_0/R_{\rm 0,i}$ with $R_{\rm 0, i}$ the Sedov-Taylor radius when the black hole forms. Numerically solving this equation for $\xi_{\rm s}(\tau_0)$ gives the location of the sonic point within the flow, and one could use this time-dependent location as that at which the fluid velocity equals zero to solve the linearized fluid equations once the black hole forms. Given the fact that the inflow boundary condition causes the shock to stall noticeably sooner compared to the simulations with a reflecting inner boundary condition (see Figure \ref{fig:setup comparison}), we expect the shock to stall and fall back to the origin in a dimensionless timescale that is comparable to the time taken for the rarefaction wave to reach the shock, i.e., Equation \eqref{sc time} evaluated at $\xi_{\rm s} = 1$. We defer a more detailed calculation of this scenario to future work. 

\subsection{Comparison of model predictions to FXTs}
\label{sec: FXTs}
From a sub-energetic explosion there are three distinct outcomes:
\begin{enumerate}
    \item The shock has enough energy to {just} reach the surface of the star, resulting in a low-energy supernova explosion. 
    \item The shock stalls before propagating through a substantial portion of the ambient medium and will be forced back to smaller radii, resulting in a failed supernova and the formation of a black hole (e.g. \citealt{Adams17}).
    \item The shock stalls in a region that is sufficiently near the stellar surface such that the edge of the star -- where the density drops precipitously -- passes through the stalled shock {before} the shock falls back to the origin. 
\end{enumerate}
Here we focus on the observational implications of the third of these three outcomes, as it may manifest as a unique astrophysical transient.  
\begin{figure*} 
 \includegraphics[width=0.495\textwidth]{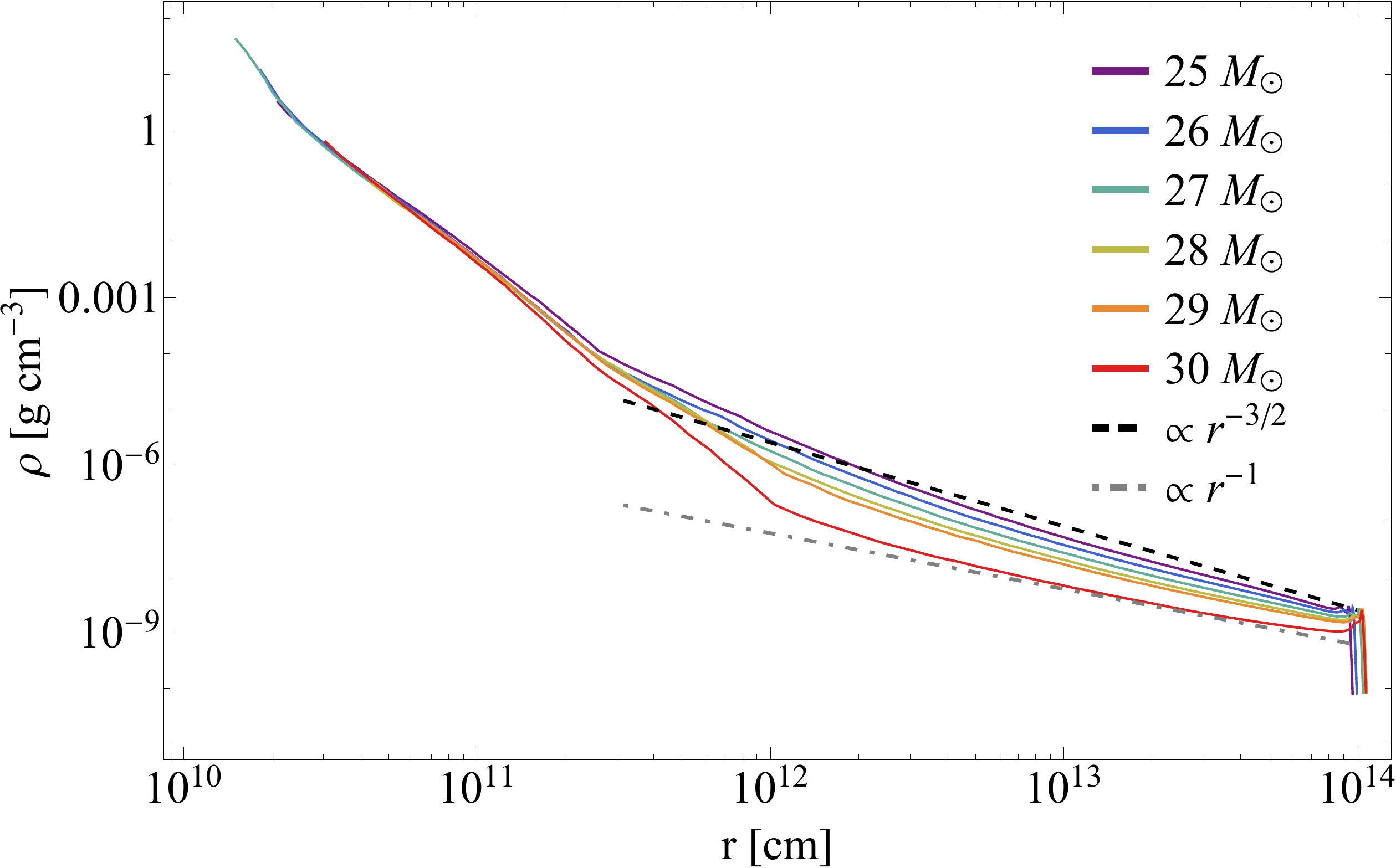}
    \includegraphics[width=0.495\textwidth]{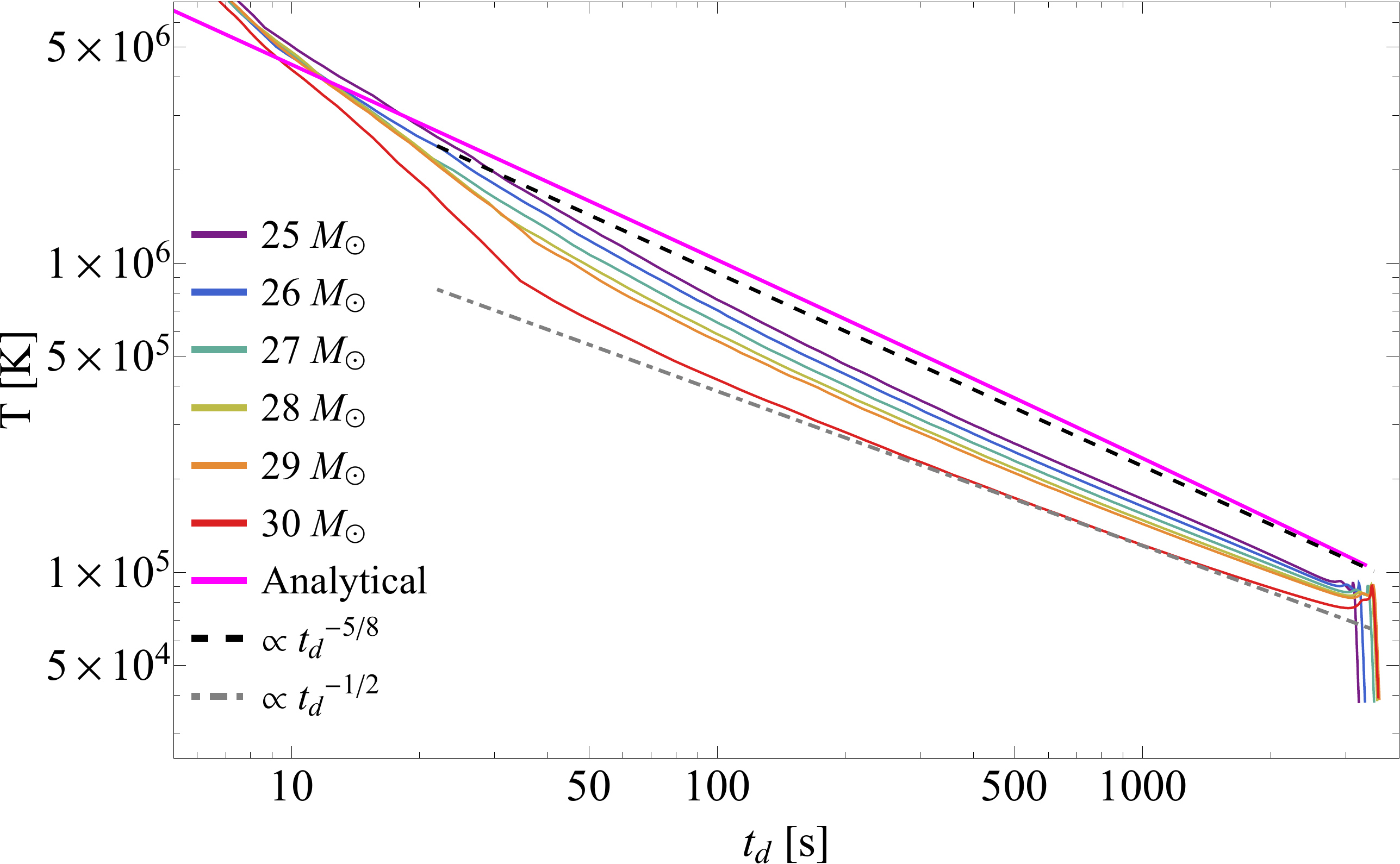}
\caption{Left: The density profiles of a set of core-collapse progenitors, taken from \citet{sukhbold16}, alongside power-law scalings. Right: The approximate temperature associated with the burst of emission that follows the stellar envelope falling through the stalled shock, as a function of the light crossing time $t_{\rm d}$, which is $R_{\rm s}/c$ with $R_{\rm s}$ the radius of the stalled shock and $c$ is the speed of light. The analytical curve is calculated by assuming an $r^{-3/2}$ scaling for the density, and using Equation \eqref{stellar pressure} to solve for the temperature. The $\propto t_{\rm d}^{-5/8}$ ($\propto t_{\rm d}^{-1/2}$) accurately reproduces the relationship between $T$ and $t_{\rm d}$ when the density profile satisfies $\rho \propto r^{-3/2}$ ($\rho \propto r^{-1}$) 
and radiation pressure dominates the equation of state, and accurately reproduces the numerically obtained (i.e., including gas pressure) result.}
\label{fig:temp}
\end{figure*}

In the event that the shock stalls after propagating through a substantial portion of the hydrogen envelope, and as long as the infalling envelope reaches the stalled shock on a timescale that is shorter than $\sim$ the dynamical time at the location of the shock (roughly the timescale over which we expect the shock to fall back to the origin), then the edge of the envelope will pass through the shock. Once the optical depth to the edge of the star is $\sim 1$, the radiation from the post-shock gas is be able to escape to the observer. Depending on the temperature of the post-shock gas, which itself depends on the radius in the star where the shock stalls, this radiation could emerge in the form of X-rays, with an initial rise time that is the longer of the diffusion time across the shock and the light-crossing time over the radius of the star (i.e., the time taken for the $\sim$ $10^{12}$ cm-sized emitting surface to ``turn on'' according to a distant observer; \citealt{Ensman92, Fernandez18}). 

To estimate the observable properties of such an event, we take values of the $9.5$\,\(M_\odot\) progenitor at core-collapse from \citet{sukhbold16}, where the density begins to be well approximated by an $r^{-3/2}$ power-law at a radius $R_{\rm i}\sim 10^{11}$ cm and the density at that point is $\sim 6\times 10^{-4} \rm g~cm^{-3}$. Using our methods from the previous sections, the density at a given stalled shock $R_{\rm s}$ is 
\begin{equation}
    \rho\left(R_{\rm s}\right)=6\times 10^{-4} \left(\frac{R_{\rm s}}{10^{11}}\right)^{-3/2}. \label{FXT Density}
\end{equation}
If the shock velocity is small relative to the freefall speed, then 
the post-shock pressure is (from Equation \ref{p bc} with $V = 0$)
\begin{equation}
    p(R_{\rm s}) \simeq \frac{2GM\rho}{R_{\rm s}}, \label{pstall}
\end{equation}
where $M$ is the mass of the star interior to the shock and, once the shock propagates to these large radii, is comparable to the mass of the entire star, and we ignored the factor of $2/(\gamma+1) \simeq 1$ for simplicity. If the pressure is a sum of gas and radiation pressure and the gas is sufficiently optically thick to thermalize the photons, then the pressure is related to the temperature via 
\begin{equation}
    p\left(R_{\rm s}\right)=\frac{\rho k T}{\mu m_{\rm H}}+\frac{1}{3}aT^4, \label{stellar pressure}
\end{equation}
where $k = 1.38\times10^{-16}$ [cgs] is Boltzmann's constant, $m_{\rm H} = 1.67\times10^{-24}$ g is the proton mass, $a = 7.56\times10^{-15}$ [cgs] is the radiation constant, and $\mu$ is the mean molecular weight. 
For a given density profile of a massive-star progenitor, we can solve Equation \eqref{stellar pressure} alongside Equation \eqref{pstall} for the temperature as a function of the radius in the star where the shock stalls. If this radius also coincides with where the optical depth to the surface (corrected for the velocity of the infalling gas; \citealt{sapir11}) is of the order unity, then we can estimate the brightening timescale by calculating the light-crossing time at that radius, which tends to be longer than the diffusion time for supergiants (whereas the diffusion time can dominate for very compact stars, such as Wolf-Rayets; \citealt{Fernandez18}). 

The left panel of Figure \ref{fig:temp} shows a set of density profiles from massive stars evolved to core collapse from the simulations in \citet{sukhbold16}, where the mass of the star is indicated in the legend. The right panel gives the temperature associated with the emission as a function of the brightening/duration timescale, $t_{\rm d} = R_{\rm s}/c$, being the light-crossing time over the shock radius. Here the analytical curve is determined by solving Equations \eqref{FXT Density}-\eqref{stellar pressure} for a given value of $R_{\rm s}$, which ranges from $10^{11}-10^{14}$ cm. As the radius at which the shock stalls decreases, the temperature increases as a consequence of both the increasing density and the increasing freefall speed, as both of these result in a larger post-shock pressure (Equation \ref{pstall}). The light-crossing time also declines, thus showing that there is a strong correlation between these two quantities. In particular, for an $r^{-3/2}$ power-law density profile, ignoring the contribution of gas pressure in Equation \eqref{stellar pressure} yields the relation $T \propto t_{\rm d}^{-5/8}$, which accurately reproduces the analytical result (which incorporates the contribution from gas pressure) and those obtained from the stellar progenitors in \citet{sukhbold16}. In the case where the density profile of the envelope is $\rho \propto r^{-n}$, we would predict (ignoring gas pressure) $T \propto t_{\rm d}^{-\left(n+1\right)/4}$. 

The observational signatures shown in Figure \ref{fig:temp} that would be associated with such a ``phantom shock breakout,'' in which the post-shock emission would be capable of emerging from the stellar surface but there would not be an associated strong explosion, are broadly consistent with the newly detected class of fast X-ray transients (FXTs; \citealt{Jonker13}). FXTs are extragalactic and non-repeating X-ray outbursts in the $\sim 0.3-10$\,keV range that last for minutes to hours \citep{Vasquez2023}. We see from the right panel of Figure \ref{fig:temp} that the range of the duration timescales predicted from this model is consistent with those of FXTs. Furthermore, for a given temperature, the radiation energy corresponding to the peak radiation frequency can be estimated by treating the progenitor as a perfect blackbody, i.e., the energy is $\sim kT$ with $T$ the temperature. 
From Figure \ref{fig:temp}, the corresponding peak energies range from $\sim 0.02-0.2$\,keV, which is roughly consistent with, but somewhat lower than, the photon energies associated with FXTs. However, our treatment of the radiation hydrodynamics is very simplistic, and a more rigorous analysis may yield somewhat higher energies that arise from repeated scatterings between the infalling and pre-shock envelope and the shocked fluid, and the eventual acceleration of the shock as it reaches a sufficiently steep density gradient.

30 FXTs have been detected serendipitously or archivally \citep{Alp20, Vasquez2022, Vasquez2023}, with the majority being the latter, and there has only been one event with a confirmed multi-wavelength counterpart after the initial detection, which is XRT 080109/SN 2008D \citep{Soderberg08}. However, observations of XRT 141001 from deep optical imaging by the Very Large Telescope were made $\sim 80$ minutes after the initial outburst, and no optical counterpart was detected \citep{Bauer17}. Due to the lack of multi-wavelength counterparts, discerning the energetics and distances to these sources, as well as their physical origin, is not straightforward. 

The progenitor(s) of FXTs is therefore unclear, and there are currently four scenarios that have been considered: core-collapse supernova shock breakout\footnote{Note that the generation of an FXT through core-collapse supernovae SBO is not the same mechanism that we propose here, as the former assumes that the SBO, and subsequent X-ray emission, is generated through a successful supernova and the ejection of material \citep{Waxman17}. When a shock breaks out of the stellar surface in a successful supernova, it is expected to be followed by bright UV/optical emission as the envelope expands and cools \citep{Falk78, Klein78, Soderberg08}, whereas we would expect an accompanying optical/UV signature to be absent if the shock stalls within the progenitor.} (SBO), X-ray binaries, off-axis gamma-ray bursts, and tidal disruption events involving an intermediate-mass black hole and a white dwarf. None of these models is particularly favored over another. However, \citet{Vasquez2023} provided an estimate of $\sim2\times10^3 - 4.5\times10^4~\rm Gpc^{-3}~yr^{-1}$ for the volumetric density rate of FXTs (see their Section 5.3), and they noted that this rate is in close agreement with the rate of blue supergiant (BSG) progenitor core-collapse supernovae ($\sim2\times10^3~\rm Gpc^{-3}~yr^{-1}$). BSGs are massive, compact ($R_{\star}\sim10^{12}$ cm) stars that have a relatively high likelihood of failing (\citealt{Fernandez18} and references therein), and thus we might expect the rate of failed explosions of these stars to be comparable to (or even exceed) the rate of successful events. This finding therefore suggests that FXTs could be associated with failed explosions, and the  ``phantom shock breakout,'' as outlined here. 

\subsection{Caveats and approximations and future directions}
\label{sec:caveats}
In our analytical and series-based approach that led to the condition on the stall radius (Section \ref{sec:Analysis}), which then showed exceptionally good agreement with numerical hydrodynamics simulations (Section \ref{sec:simulations}), the ambient medium is assumed to be in time-steady and pressureless freefall. The time-steady nature of the ambient solution is only self-consistent (i.e., solves the fluid equations) if the density profile decays with radius as $\propto r^{-3/2}$, and hence this was a necessary assumption in our model. 

As shown by the left panel of Figure \ref{fig:temp}, realistic supergiant progenitors have hydrogen envelopes that can, in some cases, be consistent with a power-law density profile that is close to $\propto r^{-3/2}$. For these stars, the assumption of time-steady freefall onto the shock is a good approximation, provided that the shock is at sufficiently small radii and the rarefaction wave resulting in the infall of successive shells of material (i.e., the mechanism responsible for initiating the collapse of the star) has reached the outer envelope of the star.

In other cases, however, the density profile is substantially flatter in the hydrogen envelope, or noticeably steeper at smaller radii. When the density profile is steeper than $\propto r^{-2}$, the shock velocity appropriate to the Sedov-Taylor solution declines less rapidly than the freefall speed, implying that there is no corresponding condition analogous to Equation \eqref{Rsofei} that will yield the radius at which the shock stalls. With that said, it may be that there is a minimum velocity (relative to the freefall speed) that the shock must have in order to transition to the strong and energy-conserving regime. An analogous situation occurs in weak shocks generated by the loss of mass during the neutron star formation, where there is a critical Mach number (which is a function of the power-law index of the ambient medium, the adiabatic index of the gas, and the mass lost to neutrinos) that differentiates between shocks that asymptotically become strong and those that transition to a weak shock with a constant Mach number near unity \citep{CQR1, CQR2, CQR3, coughlin23}. An interesting question, which we intend to analyze in future work, is whether there is an analogous condition when the ambient medium is in freefall. 

When the density profile is shallower than $\propto r^{-2}$ and the shock is at sufficiently small radii, the radius at which the shock stalls can be predicted -- at the order of magnitude level -- by assuming that the infalling gas is characterized by an analytic solution following the passage of a rarefaction wave \citep{CQR2}. It should be possible to use the more general and self-similar expressions for the rarefaction wave solutions, or those that account for the mass lost to neutrinos (either in the weak- or stronger-shock limit; \citealt{coughlin23}), for the properties of the ambient gas in the shock jump conditions, and thereby generalize the approach taken here to understand the eventual stall of the shock. This approach would also account for the finite binding energy of the material, which will have important consequences for the dynamics when the power-law index of the ambient medium is less than 2 (as in this case the binding energy diverges at large radii if the power-law profile extends indefinitely). 

We also assumed -- for simplicity, concreteness, and because it is likely accurate when the shock is still at small radii -- that the fluid is characterized by an adiabatic equation of state with an adiabatic index of $\gamma = 4/3$. Simply adopting a different adiabatic index, but otherwise performing the same analysis, is straightforward, and we would expect a stiffer (softer) equation of state to result in a larger (smaller) stall radius, i.e., if $\gamma$ is larger it would naively enable the shock to reach larger radii owing to the greater pressure support of the post-shock gas. Non-ideal effects such as nuclear dissociation are also likely important, but in general we expect such effects to increase the minimum successful explosion energy, i.e., the limits provided in Sections \ref{sec:estimates} and \ref{sec:Analysis} (specifically Equations \ref{energyapprox} and \ref{min energy}) are still valid as relatively conservative lower limits. 

Radiative effects will also modify the propagation of the shock and the dynamical evolution of the post-shock fluid in ways that are not incorporated in our analysis. In the Eddington approximation (i.e., when the photon mean-free-path can be approximated as zero compared to the fluid scale), the radiation and gas pressure linearly combine and can be treated as independent contributions to the total pressure, which -- even when the scatterers and the radiation field are considered to be in local thermodynamic equilibrium, as in Equation \eqref{stellar pressure} -- destroys the self-similarity of the problem and renders the Sedov-Taylor solution (i.e., the leading-order solution in our series approach to modeling gravitational effects) inapplicable. However, at small radii we expect the temperature to be sufficiently large that the gas can be approximated as ultra-relativistic, while at large radii the ambient density is sufficiently low that the gas pressure contributes at the $\lesssim 10\%$ level (see the discussion around Figure \ref{fig:temp} in the preceding subsection), and in both cases the $\gamma = 4/3$ equation of state is approximately valid. Finite mean-free-path effects are of critical importance as the shock nears the surface and the optical depth (from the shock to the surface) approaches unity, and investigating these effects -- and in particular the ``phantom shock breakout'' signature discussed in the previous subsection -- is an area we intend to analyze. The finite mean-free-path also plays a fundamental role in mediating the deceleration of the flow through the shock, as the propagation of photons upstream transfers outward momentum to the infalling gas (i.e., it is a photon-mediated shock owing to the fact that the photon mean-free-path will generally be much larger than the interparticle mean-free-path; \citealt{mihalas84} and references therein). Nonetheless, provided that the mean-free-path is small relative to the shock scale, treating the shock as a discontinuity that is bounded by adiabatic flow on either side is a good approximation.

Finally, in our model the central mass (responsible for the gravitational field and the infall of the gas at larger radii) was assumed to be dominated by the neutron star. While this is certainly valid at sufficiently large radii, there will come a point in a realistic star when the enclosed mass is comparable to -- and then exceeds -- this mass. It is possible to account for the self-gravity of the gas in the series-expansion approach here, but we expect this to be non-trivial, owing to the change not just in the post-shock fluid but also in the properties of the ambient gas (i.e., even pressureless freefall is complicated by the spatially variable self-gravitational field). In general, we only expect the inclusion of self-gravity to increase the lower limit on the minimum successful explosion energy.
\section{Summary and Conclusions}
\label{sec:conclusion}
When the shockwave generated through the core-collapse of a massive star is sub-energetic, such that the energy associated with the shock is akin to the binding energy of the star, the shock decelerates and can eventually propagate at a speed comparable to the escape speed of the progenitor. Therefore, even if the shock is initially strong, if it decelerates to speeds comparable to the escape speed before reaching the surface of the star, the shock can succumb to the influence of gravity and stall within the progenitor, resulting in a failed supernova. 
After providing back-of-the-envelope estimates in Section \ref{sec:estimates}, in Section \ref{sec:Analysis} we developed an analytical model for a shockwave propagating into an adiabatic, power-law ambient medium undergoing gravitational collapse with a density profile $\rho\propto r^{-3/2}$ -- which is a reasonable approximation for the outer envelope of some massive stars -- and adiabatic index $\gamma=4/3$. In particular, we analyzed perturbations to the Sedov-Taylor solution, where the perturbative (``smallness'') parameter is the ratio of the free-fall speed to the shock speed. With this approach, we showed that if the energy associated with the initial explosion is below a critical value, there exists a point where the shock stalls within the progenitor (see Figures \ref{fig:VV0 Plot} \& \ref{fig:0p3 plots}). This model therefore establishes a minimum-energy condition for a successful supernova explosion -- which is Equation \eqref{min energy} -- and we showed that for representative values of a supergiant progenitor, this energy is comparable to the binding energy of the hydrogen envelope in a red supergiant. Therefore, 
if the energy of the blastwave is below this critical value, the shock will stall before reaching the surface of the progenitor and will be unable to unbind the stellar envelope. 

In Section~\ref{sec:simulations} we compared our analytical results to those obtained with the numerical hydrodynamics code {\sc FLASH}. We showed that our analytical solution is able to accurately predict the position and velocity of the shock, as well as the post-shock velocity, density, and pressure, until the shock stalls. We showed the results of two simulations with different values of $\epsilon_{\rm i}$, which is the initial value of the free-fall speed to the shock speed and is directly related to the energy of the explosion through Equation~\eqref{enint}. The analytically predicted shock position, post-shock velocity, and post-shock density are nearly indistinguishable from their numerically simulated values, while the analytical post-shock pressure very nearly matches the numerical simulation at early times, but begins to deviate from the numerical values at small radii over time. Figure \ref{fig:0p3 plots} shows that our analytical prediction is able to very accurately predict the temporal evolution of the shock position and velocity until the shock stalls. We also compared our numerical results from the $\epsilon_{\rm i}=0.3$ simulation to the analytical settling solution, which describes the velocity, density, and pressure profiles of the fluid interior to a standing accretion shock where the fluid ``settles'' onto the surface of the neutron star. We show that the numerical simulation \emph{attempts} to match onto the settling solution, but the pressure near the point mass does not increase rapidly enough to allow the fluid to stall indefinitely. Instead, the shock continues to move inward as more mass is concentrated near the neutron star, implying that a black hole will form around the time that the shock stalls. The implications of black hole formation on the propagation of the shock were considered in Section \ref{sec:BH formation}. 

If the shock energy is only marginally below the necessary value to create a successful explosion, the shock will stall at a location that is close to the stellar surface, and it may be possible for the surface of the star to pass through the stalled shock. In this case, one would expect the passage of the stellar surface through the shock to result in the sudden emergence of radiation from the shock-heated gas, i.e., a shock-breakout-like signal but without a supernova that we denote a ``phantom shock breakout.'' To estimate the observable properties associated with this scenario, in Section \ref{sec: FXTs} we used the density profiles of six different massive star progenitors evolved to core-collapse from \citet{sukhbold16} (see Figure \ref{fig:temp}) to determine the temperature of the gas at a given stall radius  
(see Equations \ref{pstall} -- \ref{stellar pressure}). We also estimated the duration of the breakout signal by computing the light-crossing time over the radius of the stalled shock. The emission temperature as a function of the duration is then shown in the right panel of Figure \ref{fig:temp}, and is approximately given by $T \propto t_{\rm d}^{-5/8}$ when the density profile is $\rho \propto r^{-5/3}$, or $T \propto t_{\rm d}^{-\left(n+1\right)/4}$ when the density satisfies $\rho \propto r^{-n}$. From this model we recover temperatures and durations that are roughly consistent with fast X-ray transients (FXTs), which are bursts of $\sim 0.3-10$ keV emission that do not yet have an accepted origin. 

We assumed that the explosion and the ambient medium were spherically symmetric and non-rotating. When the shock is strong and well-described by the Sedov-Taylor solution, asymmetries in the initial explosion geometry and imparted to the flow from asymmetries in the ambient medium are stable, meaning that they decay with time\footnote{Unless the ambient medium is sufficiently steep that the Sedov-Taylor solution ends in a contact discontinuity, in which case the flow can be convectively unstable; \cite{goodman90}.} when modeled as perturbations on top of the Sedov-Taylor solution \citep{Ryu87}. However, it has been suggested that standing and time-steady shocks can be weakly unstable to large-angle (spherical-harmonic $\ell=1,2$) perturbations, potentially leading to an asymmetric explosion (e.g., \citealt{houck92, Foglizzo02, Blondin03, fernandez09, marek09, burrows12, dunham23}). It may therefore be the case that, as the initially strong explosion decelerates and nears the stall radius, it becomes more susceptible to angular perturbations in the ambient gas (induced by, e.g., convection in the outer envelope of the massive star) that result in the shock becoming increasingly asymmetric. It has also been suggested that even when the core-collapse does not drive an energetic explosion and a black hole is formed, the angular momentum in the convective regions of the progenitor can lead to the formation of an accretion disk and possible generation of observable outflows (see \citealt{Gilkis16, Quataert19, Antoni22, Antoni23, Soker23}). One could quantitatively investigate the temporal evolution of these asymmetries by performing a perturbation analysis with the analytic (series) solution presented here as the background state; we intend to perform such an analysis in future work. 

Finally, our analytical model only considered the case where the density profile of the ambient medium satisfies $\rho \propto r^{-3/2}$ (though see Section \ref{sec:estimates}, and particularly Equation \ref{Egen}, for an estimate of the minimum explosion energy when the ambient medium is of the form $\rho \propto r^{-n}$ at large radii), as this power-law profile results in time-steady infall of the overlying envelope. However, the density profile of a supergiant envelope is more complicated than a single power-law (see Figure \ref{fig:temp}), and the infall of material occurs in a time-dependent manner as successive shells of gas are informed of the loss of pressure support in the core. Nonetheless, it should be possible to expand upon the model considered here by using the rarefaction-wave solutions presented in \citet{coughlin19} (or the solutions that account for the neutrino-induced mass loss given in \citealt{coughlin23}) for the ambient medium, which are valid for an arbitrary power-law density profile. We plan to consider such an extension to our model in future work.

\begin{acknowledgments}
\noindent
We thank the anonymous referee for useful comments regarding the importance of asymmetries and radiative effects on the shock propagation. DAP acknowledges support from the Summer Pre-Dissertation Fellowship through the Graduate School at Syracuse University. DAP also thanks Suman Kumar Kundu for providing guidance and useful discussions. ERC acknowledges support from the National Science Foundation through grant AST-2006684 and the Oakridge Associated Universities through a Ralph E.~Powe Junior Faculty Enhancement award.
\end{acknowledgments}

\newpage
\appendix
\section{Fluid Equations}
\label{sec:appendix}
Here we write the dimensionless continuity, radial momentum, and entropy equations up to the second order in the ratio of the freefall speed to the shock speed. The zeroth-order equations yield the well-known Sedov-Taylor solution, and are given by
\begin{align}
    \label{zeroth cont}
    -\frac{3}{2}g_0-\xi \frac{\partial{g_0}}{\partial{\xi}}+\frac{1}{\xi^2} \frac{\partial{}}{\partial{\xi}}\left[\xi^2 g_0 f_0\right]&=0, \\
    \label{zeroth momentum}
    -\frac{3}{4}f_0 + \left(f_0 - \xi\right)\frac{\partial{f_0}}{\partial{\xi}}+\frac{1}{g_0}\frac{\partial{}}{\partial{\xi}}h_0&=0, \\
    \label{zeroth entropy}
    -\frac{3}{2}+\frac{3}{2}\left(\gamma-1\right)+\left(f_0-\xi\right)\frac{\partial{}}{\partial{\xi}}s_0&=0.
\end{align}
The first order equations are 
\begin{align}
    \label{first cont}
    -\frac{5}{4}g_1-\xi \frac{\partial{}}{\partial{\xi}}g_1+\frac{1}{\xi^2} \frac{\partial{}}{\partial{\xi}}\left[\xi^2\left(f_0 g_1 + g_0 f_1\right)\right]&=0, \\
    \label{first momentum}
    -\frac{1}{2}f_1+\frac{1}{2}\alpha_1 f_0 - \xi\frac{\partial{f_1}}{\partial{\xi}}+\frac{\partial{}}{\partial{\xi}}\left[f_0 f_1\right]+\frac{1}{g_0}\left(\frac{\partial{h_1}}{\partial{\xi}}-\frac{g_1}{g_0}\frac{\partial{h_0}}{\partial{\xi}}\right)&=0, \\
    \label{first entropy}
    \alpha_1 + \frac{1}{4}s_1+\left(f_0 - \xi\right)\frac{\partial{}}{\partial{\xi}}s_1+f_1\frac{\partial{}}{\partial{\xi}}s_0&=0,
\end{align}
and the second order equations are
\begin{align}
    \label{second cont}
    -g_2-\frac{1}{16}\alpha_1g_1-\xi \frac{\partial{g_2}}{\partial{\xi}}+\frac{1}{\xi^2} \frac{\partial{}}{\partial{\xi}}\left[\xi^2\left(f_0 g_2 + f_1 g_1+ g_0 f_2\right)\right]&=0, \\
    \label{second momentum}
    \begin{split}
    -\frac{3}{4}\left(f_2-\frac{2}{3}\alpha_1 f_1+\frac{3}{16}\left(5{\alpha_1}^2-8\alpha_2\right)f_0\right)-\frac{1}{16}\alpha_1 f_1 +\frac{1}{2}f_2 - \xi\frac{\partial{f_2}}{\partial{\xi}}+\left(f_0\frac{\partial{f_2}}{\partial{\xi}}+f_1\frac{\partial{f_1}}{\partial{\xi}}+f_2\frac{\partial{f_0}}{\partial{\xi}}\right)
    \\
    +\frac{1}{g_0}\left(\frac{\partial{h_2}}{\partial{\xi}}-\frac{g_1}{g_0}\frac{\partial{h_1}}{\partial{\xi}}+\left(\frac{{g_1}^2}{{g_0}^2}-\frac{g_2}{g_0}\right)\frac{\partial{h_0}}{\partial{\xi}}\right)&=-\frac{1}{2{\xi}^2},
    \end{split} \\
    \label{second entropy}
    -\frac{9}{32}\left(5{\alpha_1}^2-8\alpha_2\right)-\frac{1}{16}\alpha_1 s_1 +\frac{1}{2}s_2+\left(f_0 - \xi\right)\frac{\partial{s_2}}{\partial{\xi}}+f_1\frac{\partial{s_1}}{\partial{\xi}}+f_2\frac{\partial{s_0}}{\partial{\xi}}&=0.
\end{align}
where
\begin{equation}
\begin{split}
    s&=\ln\left(\frac{h}{g^{\gamma}}\right)=\ln\left(\frac{h_0}{g_0^{\gamma}}\right)+\epsilon_0\left(\frac{h_1}{h_0}-\frac{\gamma g_1}{g_0}\right)+
    {\epsilon_0}^2\left(\frac{h_2}{h_0}-\frac{\gamma h_1 g_1}{h_0 g_0} +\frac{1}{2}\gamma\left(\gamma+1\right)\frac{{g_1}^2}{{g_0}^2}-\frac{\gamma g_2}{g_0}-\frac{1}{2}{\left(\frac{h_1}{h_0}-\frac{\gamma g_1}{g_0}\right)}^2\right)
    \\
    &\equiv s_0 + \epsilon_0 s_1 + {\epsilon_0}^2 s_2.
    \end{split}
\end{equation}
Each of these systems of differential equations can be solved numerically by integrating inward from the location of the shock ($\xi=1$), at which the functions satisfy the boundary conditions
\begin{align}
\nonumber
    & f_0\left(1\right)=\frac{2}{\gamma+1},\quad f_1\left(1\right)=-\frac{\gamma-1}{\gamma+1},\quad f_2\left(1\right)=\frac{7}{4}\frac{\gamma-1}{\gamma+1}\alpha_1, \\
    & g_0\left(1\right)=\frac{\gamma+1}{\gamma-1},\quad g_1\left(1\right)=g_2\left(1\right)=0, \\
\nonumber
    & h_0\left(1\right)=\frac{2}{\gamma+1},\quad h_1\left(1\right)=\frac{4}{\gamma+1},\quad h_2\left(1\right)=\frac{2}{\gamma+1}\left(1-\frac{7}{2}\alpha_1\right).
\end{align}
The $\alpha_{\rm n}$ values, which arise from the zero-velocity boundary condition at the origin, are given in Table~\ref{tab:eigenvalues}.

\bibliographystyle{aasjournal}
\bibliography{ref}

\end{document}